\documentclass[useAMS,usenatbib]{mnras}

\usepackage{graphicx}
\usepackage{ulem}
\usepackage[table]{xcolor}
\usepackage{amsmath}
\usepackage{caption}
\usepackage[T1]{fontenc}
\usepackage{bold-extra}

\def\ltsima{$\; \buildrel < \over \sim \;$}
\def\simlt{\lower.5ex\hbox{\ltsima}}
\def\gtsima{$\; \buildrel > \over \sim \;$}
\def\simgt{\lower.5ex\hbox{\gtsima}}
\DeclareFontFamily{U}{mathx}{\hyphenchar\font45}
\DeclareFontShape{U}{mathx}{m}{n}{<-> mathx10}{}
\DeclareSymbolFont{mathx}{U}{mathx}{m}{n}
\DeclareMathAccent{\widebar}{0}{mathx}{"73}

\def\siglos{\sigma_{\text{LOS}}}

\def\betastar{\beta^*}
\def\dd{\text{d}}

\def\GravSphere{{\sc GravSphere}}
\def\GravSpherebf{{\bfseries\scshape GravSphere}}
\def\pkdgrav{{\sc PkdGRAV-2}}

\def\coreNFW{{\sc coreNFW}}
\def\coreNFWtides{{\sc coreNFWtides}}

\def\EMCEE{{\sc emcee}}

\def\betastar{\tilde{\beta}}
\def\vsone{v_{s1}}
\def\vstwo{v_{s2}}
\def\vlosfour{\langle v_{\rm LOS}^4 \rangle}

\usepackage{array}
\newcolumntype{L}[1]{>{\raggedright\let\newline\\\arraybackslash\hspace{0pt}}m{#1}}
\newcolumntype{C}[1]{>{\centering\let\newline\\\arraybackslash\hspace{0pt}}m{#1}}
\newcolumntype{R}[1]{>{\raggedleft\let\newline\\\arraybackslash\hspace{0pt}}m{#1}}

\title[A dark matter cusp in Draco]{The case for a cold dark matter cusp in Draco}

\author[Read]{J. I. Read$^1$\thanks{E-mail: justin.inglis.read@gmail.com}, M. G. Walker$^2$, P. Steger$^3$\\
  $^1$Department of Physics, University of Surrey, Guildford, GU2 7XH, UK\\
  $^2$McWilliams Center for Cosmology, Department of Physics, Carnegie Mellon University, 5000 Forbes Ave., \\
  Pittsburgh, PA 15213, United States\\
  $^3$Institute for Astronomy, Department of Physics, ETH Z\"urich, Wolfgang-Pauli-Strasse 27, CH-8093 Z\"urich, Switzerland\\
}

\begin{document}

\maketitle

\begin{abstract}
We use a new mass modelling method, \GravSphere, to measure the central dark matter density profile of the Draco dwarf spheroidal galaxy. Draco's star formation shut down long ago, making it a prime candidate for hosting a `pristine' dark matter cusp, unaffected by stellar feedback during galaxy formation. We first test \GravSphere\ on a suite of tidally stripped mock `Draco'-like dwarfs. We show that we are able to correctly infer the dark matter density profile of both cusped and cored mocks within our 95\% confidence intervals. While we obtain only a weak inference on the logarithmic slope of these density profiles, we are able to obtain a robust inference of the amplitude of the inner dark matter density at 150\,pc, $\rho_{\rm DM}(150\,{\rm pc})$. We show that, combined with constraints on the density profile at larger radii, this is sufficient to distinguish a $\Lambda$ Cold Dark Matter ($\Lambda$CDM) cusp -- that has $\rho_{\rm DM}(150\,{\rm pc}) \simgt 1.8 \times 10^8\,{\rm M}_\odot \,{\rm kpc}^{-3}$ -- from alternative dark matter models that have lower inner densities. We then apply \GravSphere\ to the real Draco data. We find that Draco has an inner dark matter density of $\rho_{\rm DM}(150\,{\rm pc}) =  2.4_{-0.6}^{+0.5} \times 10^8\,{\rm M}_\odot \,{\rm kpc}^{-3}$, consistent with a $\Lambda$CDM cusp. Using a velocity independent SIDM model, calibrated on $\Lambda$SIDM cosmological simulations, we show that Draco's high central density gives an upper bound on the SIDM cross section of $\sigma/m < 0.57$\,cm$^2$\,g$^{-1}$ at 99\% confidence. We conclude that the inner density of nearby dwarf galaxies like Draco provides a new and competitive probe of dark matter models.
\end{abstract}


\section{Introduction}\label{sec:intro}

The standard $\Lambda$CDM cosmological gives an excellent description of the cosmic microwave background radiation \citep[e.g.][]{2014A&A...571A..16P}, the growth of structure on large scales \citep[e.g.][]{2006Natur.440.1137S,2016JCAP...08..012B} and the offsets between mass and light in weak lensing systems \citep[e.g.][]{2006ApJ...648L.109C,2015Sci...347.1462H}. Yet, it contains two mysterious ingredients -- dark matter and dark energy -- that remain elusive. One path to constraining the nature of dark matter is to probe its distribution on ever smaller scales, where $\Lambda$CDM is less well-tested and where differences between competing dark matter models are maximised \citep[e.g.][]{2000PhRvL..84.3760S,2001ApJ...556...93B,2016JCAP...08..012B}. This `near-field cosmology' showed early promise, turning up a host of `small scale puzzles' that continue to challenge $\Lambda$CDM today \citep[e.g.][]{2017ARA&A..55..343B}. The oldest of these is the `cusp-core' problem: the inner rotation curves of dwarf irregular galaxies rise less steeply than expected from pure dark matter structure formation simulations \citep{1994Natur.370..629M,1994ApJ...427L...1F}. This implies that the central dark matter density of these dwarfs is lower than expected in a pure-dark matter $\Lambda$CDM cosmology.

Many solutions have been proposed to the cusp-core problem, falling in to three main classes. The first class changes the nature of dark matter itself. Such models include `Self Interacting Dark Matter' (SIDM; \citealt{2000PhRvL..84.3760S,2013MNRAS.430...81R,2015MNRAS.453...29E,2016PhRvL.116d1302K,2017MNRAS.470.1542S,2017MNRAS.472.2945R}); `Warm Dark Matter' (e.g. \citealt{2001ApJ...561...35D,2001ApJ...556...93B,2001ApJ...559..516A,2014MNRAS.439..300L,2017MNRAS.470.1542S}, but see \citealt{2012MNRAS.424.1105M} and \citealt{2013MNRAS.430.2346S}); `fuzzy DM' \citep{2000PhRvL..85.1158H}, `fluid' DM \citep{2000ApJ...534L.127P} and `wave-like' DM \citep{2014NatPh..10..496S}, to name a few. The second class invokes some problem with the data, arguing that the measurements are wrong because of poor resolution \citep[e.g.][]{2010AdAst2010E...5D}, incorrectly measured inclinations \citep[e.g.][]{2004ApJ...617.1059R,2016MNRAS.462.3628R}, disequilibrium \citep[e.g.][]{2016MNRAS.462.3628R}, unmodelled pressure support \citep[e.g.][]{2004ApJ...617.1059R,2007ApJ...657..773V,2017MNRAS.466...63P} or unmodelled triaxiality/non-circular motions \citep[e.g.][]{2004ApJ...617.1059R,2007ApJ...657..773V,2011MNRAS.414.3617K,2017arXiv170607478O}. The third class invokes missing `baryonic physics'. In this, the gravitational interaction between normal `baryonic' matter (stars and gas) -- that are not included in the pure dark matter simulations discussed above -- physically transforms the dark matter cusp to a core \citep[e.g.][]{1996MNRAS.283L..72N,2005MNRAS.356..107R,2012MNRAS.421.3464P,2001ApJ...560..636E,2009ApJ...698.2093D,2010ApJ...725.1707G,2015MNRAS.446.1820N}.

Of the three solutions, above, the first is the most exciting as it reveals to us something about the nature of dark matter. However, for this to be convincing, the other two classes must first be ruled out. Much work has gone into probing the second class of solution \citep[e.g.][]{2007ApJ...657..773V,2011MNRAS.414.3617K,2016MNRAS.462.3628R,2017MNRAS.466...63P}, typically by applying standard techniques to mock data. While individual cases can be found where the standard techniques fail, none of the potential problems discussed to date would systematically bias all dwarfs towards apparent dark matter cores. Yet, this is what is needed to explain the data \citep[e.g.][]{2017MNRAS.466...63P,2017A&A...605A..55A}.

The third solution, however, has proven more promising. \citet{1996MNRAS.283L..72N} were the first to suggest that impulsive gas blow out could irreversibly heat dark matter in dwarf galaxies. For a single burst the effect is small \citep{2002MNRAS.333..299G}. However, multiple bursts can gradually transform a cusp to a core \citep{2005MNRAS.356..107R}. Such an effect is now seen in simulations of dwarf galaxies that resolve the clumpy interstellar medium \citep[e.g.][]{2008Sci...319..174M,2012MNRAS.421.3464P,2014Natur.506..171P}. Furthermore, these simulations make several testable predictions: (i) star formation should be bursty with a duty cycle comparable to the local dynamical time, and a peak-to-trough ratio of $5-10$ \citep{2013MNRAS.429.3068T}; (ii) stars should be similarly heated along with the dark matter, leading to a `hot' stellar distribution with $v/\sigma \sim 1$, where $v$ is the rotational velocity and $\sigma$ is the velocity dispersion \citep{2005MNRAS.356..107R,2012ApJ...750...33L,2013MNRAS.429.3068T}; (iii) dark matter cores should have a size of order the projected half stellar mass radius, $R_{1/2}$ \citep{2015MNRAS.454.2092O,2016MNRAS.459.2573R}; and (iv) galaxies that stopped forming stars long ago should be cuspier than those that formed stars for a Hubble time \citep{2012ApJ...759L..42P,2014MNRAS.437..415D,2015MNRAS.454.2092O,2016MNRAS.459.2573R}. Predictions (i)-(iii) have now all been tested against real data and passed \citep{2012ApJ...750...33L,2014MNRAS.441.2717K,2016ApJ...820..131E,2017MNRAS.465.2420W,2017MNRAS.466...88S,2017MNRAS.467.2019R,2017A&A...605A..55A}. However, prediction (iv) remains elusive. The challenge is to measure the central dark matter density profile in a galaxy that is no longer star forming today. However, such galaxies are, by construction, devoid of HI gas and so we can no longer use a rotation curve to reconstruct their mass distribution. Instead, we must make use of the velocities of their individual stars. This is made difficult by the `$\rho-\beta$ degeneracy' \citep[e.g.][]{1982MNRAS.200..361B,1990AJ.....99.1548M,2017MNRAS.471.4541R}. This is a degeneracy between the radial density profile, $\rho(r)$, and the unknown orbit distribution of the stars. This latter is typically parameterised by the `velocity anisotropy parameter', $\beta(r)$ (equation \ref{eqn:beta}), that is hard to constrain with line-of-sight velocities alone \citep[e.g.][]{2017MNRAS.471.4541R}.

Several methods have been proposed to break the $\rho-\beta$ degeneracy, including modelling multiple populations of different scale lengths all moving in the same potential \citep[e.g.][]{2008ApJ...681L..13B,2011ApJ...742...20W,2016MNRAS.463.1117Z}, using higher order velocity moments \citep[e.g.][]{2009MNRAS.394L.102L}, Schwarzschild methods \citep[e.g.][]{2013MNRAS.433.3173B,2013ApJ...763...91J}, proper motions \citep{2007ApJ...657L...1S,2017arXiv171108945M}, and `Virial Shape Parameters' (VSPs; \citealt{1990AJ.....99.1548M,2014MNRAS.441.1584R,2017MNRAS.471.4541R}). In this paper, we use this latter method, implemented in the non-parametric Jeans modelling code, \GravSphere\ \citep{2017MNRAS.471.4541R}. This has the advantages that: (i) only line of sight velocity data are required; (ii) we need make no assumption about the form of the velocity distribution function; and (iii) no population splitting is required \citep{2017MNRAS.471.4541R}. We focus on modelling the dark matter distribution in the Draco dwarf spheroidal galaxy (dSph). Draco was first discovered by \citet{1955PASP...67...27W} using photographic plates. It lies just 76\,kpc from the Galactic centre and, with a stellar mass of $M_* = 0.29 \times 10^6$\,M$_\odot$, is one of the most dark matter dominated galaxies in the Universe \citep[e.g.][]{2001ApJ...563L.115K,2012AJ....144....4M}. Draco is particularly interesting because it stopped forming stars $\sim 10$\,Gyrs ago \citep{2001AJ....122.2524A}. This makes it a prime candidate for hosting a `pristine' dark matter cusp within its projected stellar half light radius, $R_{1/2} = 0.22$\,kpc \citep{2012AJ....144....4M}, unaffected by bursty star formation (e.g. \citealt{2015MNRAS.450.3920B,2016MNRAS.459.2573R,2018MNRAS.479.1514B}). However, unlike other galaxies with similarly old-age stellar populations, Draco also has a large number of $\sim 500$ member stars with well-measured radial velocities \citep{2015MNRAS.448.2717W}. We will show in \S\ref{sec:mock} that this is sufficient to break the $\rho-\beta$ degeneracy and measure the inner dark matter density, even if Draco has experienced tidal stripping by the Milky Way down to its projected half stellar mass radius, $R_{1/2}$.

This paper is organised as follows. In \S\ref{sec:cuspcore}, we briefly review the cusp-core problem in $\Lambda$CDM. We show that, while the cusp-core problem is usually framed in terms of the inner logarithmic slope of the density profile, the amplitude of the central density -- that is easier to determine observationally -- is sufficient to constrain interesting models like SIDM. In \S\ref{sec:method}, we briefly describe the \GravSphere\ method; a more complete description, including a large number of tests is given in \citet{2017MNRAS.471.4541R}. We also describe our SIDM model and its calibration on $\Lambda$SIDM simulations (\S\ref{sec:sidmmodel}). In \S\ref{sec:mock} we test \GravSphere\ on a suite of tidally stripped mock `Draco'-like dwarfs, showing that we are able to recover the dark matter density profile of these mocks within our 95\% confidence intervals. In \S\ref{sec:data}, we describe our data compilation and reduction for Draco. In \S\ref{sec:results}, we present our results from applying \GravSphere\ to these data. We show that our \GravSphere\ models for Draco favour a large central density, consistent with a dark matter cusp and we use this to to place a new constraint on the self-interaction cross section of dark matter. In \S\ref{sec:discussion}, we discuss the caveats inherent in our modelling and the implications of our results in the context of $\Lambda$CDM. Finally, in \S\ref{sec:conclusions} we present our conclusions.

\section{The cusp-core problem in $\Lambda$CDM}\label{sec:cuspcore}

Pure dark matter simulations in $\Lambda$CDM predict halos that have a density profile that is well-fit (at the $\sim 10\%$ level\footnote{An Einasto profile provides a slightly better fit \citep[e.g.][]{2006AJ....132.2685M}, though even this can be improved upon \citep[e.g.][]{2009MNRAS.398L..21S}. The classic NFW profile will suffice for our study here.}) by a split-power law known as the `Navarro, Frenk \& White' (NFW) profile \citep{1996ApJ...462..563N}: 

\begin{equation} 
\rho_{\rm NFW}(r) = \rho_0 \left(\frac{r}{r_s}\right)^{-1}\left(1 + \frac{r}{r_s}\right)^{-2}
\label{eqn:rhoNFW}
\end{equation}
where the central density $\rho_0$ and scale length $r_s$ are given by: 
\begin{equation} 
\rho_0 = \rho_{\rm crit} \Delta c_{200}^3 g_c / 3 \,\,\,\, ; \,\,\,\, r_s = r_{200} / c_{200}
\end{equation}
\begin{equation}
g_c = \frac{1}{{\rm log}\left(1+c_{200}\right)-\frac{c_{200}}{1+c_{200}}}
\end{equation}
and
\begin{equation} 
r_{200} = \left[\frac{3}{4} M_{200} \frac{1}{\pi \Delta \rho_{\rm crit}}\right]^{1/3}
\label{eqn:r200}
\end{equation} 
where $c_{200}$ is the dimensionless {\it concentration parameter}; $\Delta = 200$ is the over-density parameter; $\rho_{\rm crit} = 136.05$\,M$_\odot$\,kpc$^{-3}$ is the critical density of the Universe at redshift $z=0$; $r_{200}$ is the `virial' radius at which the mean enclosed density is $\Delta \times \rho_{\rm crit}$; and $M_{200}$ is the `virial' mass -- the mass within $r_{200}$.

The mass and concentration of halos in $\Lambda$CDM are correlated. \citet{2014MNRAS.441.3359D} find a best-fit relation for field halos of:

\begin{equation}
\log_{10}(c_{200}) = 0.905 - 0.101 \log_{10}(M_{200} h - 12)
\label{eqn:M200c200}
\end{equation}
with scatter $\Delta \log_{10}(c_{200}) = 0.1$, where $h \sim 0.7$ is the Hubble parameter. (Note that subhalos are found to be more concentrated than field halos, most likely due to tidal stripping steepening their outer density profiles, e.g. \citealt{2008MNRAS.391.1685S} and \citealt{2011ApJ...740..102K}. We will consider this further in \S\ref{sec:sidmmodel}.)

\begin{figure}
\begin{center}
\includegraphics[width=0.45\textwidth]{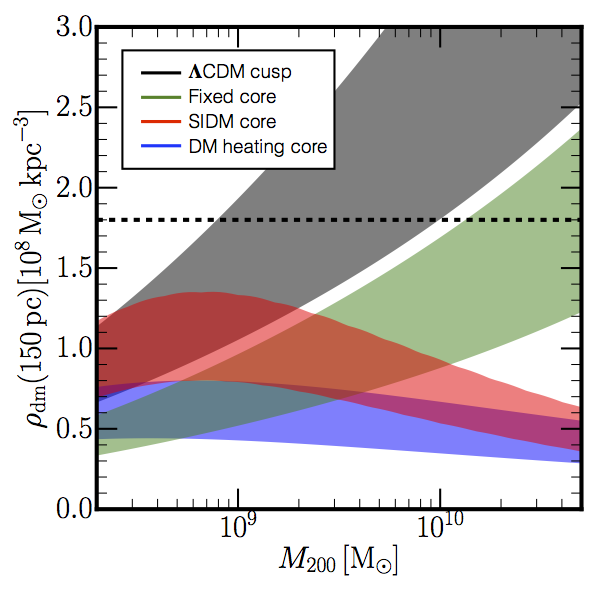}
\caption{Dark matter cusps in $\Lambda$CDM have a high central density. This plot shows the dark matter density at 150\,pc ($\rho_{\rm DM}(150\,{\rm pc})$) as a function of halo mass $M_{200}$ for four different models. The grey band is for predictions from pure dark matter simulations in a $\Lambda$CDM cosmology (i.e. NFW profiles; equation \ref{eqn:rhoNFW}). The width of the band owes to the expected $1\sigma$ scatter in the mass-concentration relation (equation \ref{eqn:M200c200}). The green band marks the same but for a \coreNFW\ profile with a fixed visible core size of $r_{cv} = 250$\,pc. The red band shows the halos in a SIDM model with a self-interaction cross section of $\sigma/m = 0.25$\,cm$^2$/g (see \S\ref{sec:sidmmodel} for details of this model). The blue band shows a model in which dark matter is `heated up' by bursty star formation, assuming that there has been sufficient star formation for core formation to complete \citep{2014MNRAS.437..415D,2015MNRAS.454.2092O,2016MNRAS.459.2573R}. Notice that, for plausible pre-infall halo masses for Draco ($10^9-10^{10}$\,M$_\odot$), a single measurement of $\rho_{\rm DM}(150\,{\rm pc}) > 1.8 \times 10^8$\,M$_\odot$\,kpc$^{-3}$ (horizontal dashed line) would imply that Draco has a visible core size $r_{cv} \simlt 250$\,pc, favouring the cusped models (grey band) over the SIDM and `dark matter heating' cored models.}
\label{fig:cuspdefine}
\end{center}
\end{figure}

While pure dark matter simulations in $\Lambda$CDM predict dense central cusps, modern simulations that include the effects of gas cooling, star formation and feedback find that -- provided sufficient star formation takes place -- these dense cusps are transformed to cores in the centres of dwarf galaxies (see \S\ref{sec:intro}). \citet{2016MNRAS.459.2573R} parameterise this transformation with the `\coreNFW' profile that has a cumulative mass given by:

\begin{equation}
M_{\rm cNFW}(<r) = M_{\rm NFW}(<r) f^n
\label{eqn:coreNFW}
\end{equation}
where $M_{\rm NFW}(<r)$ is the cumulative mass of the NFW profile: 

\begin{equation} 
M_{\rm NFW}(r) = M_{200} g_c \left[\ln\left(1+\frac{r}{r_s}\right) - \frac{r}{r_s}\left(1 + \frac{r}{r_s}\right)^{-1}\right]
\label{eqn:MNFW}
\end{equation}
and the function $f^n$ generates a shallower profile below a core-size parameter, $r_c$: 

\begin{equation} 
f^n = \left[\tanh\left(\frac{r}{r_c}\right)\right]^n
\end{equation}
The density profile of this \coreNFW\ model is given by:

\begin{equation} 
\rho_{\rm cNFW}(r) = f^n \rho_{\rm NFW} + \frac{n f^{n-1} (1-f^2)}{4\pi r^2 r_c} M_{\rm NFW}
\label{eqn:rhocNFW}
\end{equation} 
\citet{2016MNRAS.459.2573R} find that their dark matter density profiles become visibly cored below $R_{1/2}$, which corresponds to a core size parameter of $r_c = 1.75 R_{1/2}$. For this reason, we define here a `visible core size parameter', $r_{cv} \equiv r_c / 1.75$. Other groups using different simulation codes and sub-grid physics recipes have found similar results (e.g. \citealt{2014MNRAS.437..415D} and \citealt{2015MNRAS.454.2092O}). Indeed, \citet{2017MNRAS.470.1542S} and \citet{2017A&A...605A..55A} show that the two main fitting functions proposed in the literature to date -- the \coreNFW\ profile, above, and the \citealt{2014MNRAS.437..415D} profile -- produce near-identical results when applied to both simulated and real data.

In Figure \ref{fig:cuspdefine}, we show the inner dark matter density as a function of halo mass $M_{200}$ for four different halo models. We define `inner' to be 150\,{\rm pc} which is ${\sim} 0.75 R_{1/2}$ for Draco (see \S\ref{sec:data}). This is a compromise between picking a radius that is inside $R_{1/2}$, but not so small that we are not able to constrain the dark matter density observationally. The grey band shows predictions from pure dark matter simulations in a $\Lambda$CDM cosmology (i.e. NFW profiles; equation \ref{eqn:rhoNFW}). The width of the band owes to the expected $1\sigma$ scatter in the mass-concentration relation (equation \ref{eqn:M200c200}). The green band marks the same but for a \coreNFW\ profile with a fixed visible core size of $r_{cv} = 250$\,pc. The red band shows the halos in a SIDM model with a self-interaction cross section of $\sigma/m = 0.25$\,cm$^2$/g (see \S\ref{sec:sidmmodel} for details of this model). The blue band shows the expectation for models in which dark matter is `heated up' by bursty star formation, assuming complete core formation (i.e. a full Hubble time of star formation). In this case, the dark matter core size is expected to scale with the projected half light radius of the stars, $R_{1/2}$ \citep{2014MNRAS.437..415D,2015MNRAS.454.2092O,2016MNRAS.459.2573R}. Since $R_{1/2} \sim 0.015\,r_{200}$ \citep{2013ApJ...764L..31K}, this gives a visible core size of $r_{cv} = 0.015\,r_{200}$. Notice that for plausible pre-infall halo masses for Draco ($M_{200} = 10^9-10^{10}$\,M$_\odot$), a single measurement of $\rho_{\rm DM}(150\,{\rm pc}) > 1.8 \times 10^8$\,M$_\odot$\,kpc$^{-3}$ (horizontal dashed line) would imply that Draco has a visible core size $r_{cv} \simlt 250$\,pc, favouring the cusped models (grey band) over the SIDM and `dark matter heating' cored models (red, green and blue bands).

Many studies of the cusp-core problem have focussed on measuring the logarithmic slope of the density profile, $\gamma_{\rm DM}(r) \equiv d \ln \rho_{\rm DM} / d\ln r(r)$, or the asymptotic slope $\gamma_{\rm DM}(r \rightarrow 0)$. Both of these are challenging to measure, as we shall show in \S\ref{sec:mock}. However, as can be seen in Figure \ref{fig:cuspdefine}, the amplitude of the central density, combined with information on $\rho_{\rm DM}(r)$ at larger radii, is already sufficient to distinguish interesting cosmological models, for example SIDM (\S\ref{sec:sidmmodel}) or models in which a central dark matter core forms in response to stellar feedback.

\section{Method}\label{sec:method}

\subsection{\GravSpherebf}\label{sec:gravsphere}

\GravSphere\ is described and tested in detail in \citet{2017MNRAS.471.4541R}. It solves the projected spherical Jeans equation \citep{1922MNRAS..82..122J,1982MNRAS.200..361B}:

\begin{equation}
 \siglos^2(R) = \frac{2}{\Sigma(R)}\int_R^\infty \left(1\!-\!\beta\frac{R^2}{r^2}\right)
    \nu\sigma_r^2\,\frac{r\,\dd r}{\sqrt{r^2\!-\!R^2}} \ ,
    \label{eqn:LOS}
\end{equation}
where $\Sigma(R)$ denotes the tracer surface mass density at projected radius $R$; $\nu(r)$ is the spherically averaged tracer density; and $\beta(r)$ is the velocity anisotropy:

\begin{equation}
\beta = 1 - \frac{\sigma_t^2}{\sigma_r^2}
\label{eqn:beta}
\end{equation}
where $\sigma_t$ and $\sigma_r$ are the tangential and radial velocity dispersions, respectively, and $\sigma_r$ is given by \citep{1994MNRAS.270..271V,2005MNRAS.362...95M}:
\begin{equation}
\sigma_r^2(r) = \frac{1}{\nu(r) g(r)} \int_r^\infty \frac{GM(\tilde{r})\nu(\tilde{r})}{\tilde{r}^2} g(\tilde{r}) \dd \tilde{r}
\label{eqn:main}
\end{equation}
where:
\begin{equation}
g(r) = \exp\left(2\int \frac{\beta(r)}{r}\dd r\right)
\label{eqn:ffunc}
\end{equation}
and $M(r)$ is the cumulative mass of the dwarf galaxy (due to all stars, gas, dark matter etc.).

\GravSphere\ uses a free-form, or `non-parametric', model for $M(r)$ that comprises a contribution from all visible matter and a contribution from dark matter that is described by a sequence of power laws defined on a set of radial bins. In this paper, these bins are defined at $[0.25,0.5,1,2,4]R_{1/2}$ where $R_{1/2}$ is the projected half light radius of the tracer stars. The tracer light profile is also non-parametric, using a series sum of Plummer spheres, as in \citet{2016MNRAS.459.3349R}. The velocity anisotropy is given by a form that makes $g(r)$ analytic: 

\begin{equation} 
\beta(r) = \beta_0 + \left(\beta_\infty-\beta_0\right)\frac{1}{1 + \left(\frac{r_0}{r}\right)^n}
\label{eqn:betagravsphere}
\end{equation}
where $\beta_0$ is the inner asymptotic anisotropy, $\beta_\infty$ is the outer asymptotic anisotropy, $r_0$ is a transition radius, and $n$ controls the sharpness of the transition.

To avoid infinities in $\beta$ for highly tangential orbits, for model fitting we use a symmetrised $\betastar$ \citep{2006MNRAS.367..387R,2017MNRAS.471.4541R}:

\begin{equation} 
\betastar =  \frac{\sigma_r^2 - \sigma_t^2}{\sigma_r^2 + \sigma_t^2} = \frac{\beta}{2-\beta}
\label{eqn:betastar}
\end{equation} 
where $\betastar = -1$ corresponds to full tangential anisotropy; $\betastar = 1$ to full radial anisotropy; and $\betastar = 0$ to isotropy. We assume flat priors on $-1 < \betastar_{0,\infty} < 1$ such that we give equal weight to tangentially and radially anisotropic models.

By default, \GravSphere\ also fits for the two higher order `Virial Shape Parameters' (VSPs; \citealt{1990AJ.....99.1548M,2014MNRAS.441.1584R,2017MNRAS.471.4541R}):

\begin{eqnarray} 
\vsone & = & \frac{2}{5}\, \int_0^{\infty} G M\, (5-2\beta) \,\nu \sigma_r^2 \,r \,\dd r \\
\label{eqn:vs1}
& = & \int_0^{\infty} \Sigma \vlosfour\, R\, \dd R
\label{eqn:vs1data}
\end{eqnarray}
and
\begin{eqnarray} 
\vstwo & = & \frac{4}{35} \,\int_0^{\infty} G M\, (7-6\beta)\, \nu \sigma_r^2 \,r^3 \,\dd r \\
\label{eqn:vs2}
& = & \int_0^{\infty} \Sigma \vlosfour\, R^3\, \dd R \ .
\label{eqn:vs2data}
\end{eqnarray}
These VSPs involve fourth-order moments of the line-of-sight velocities $\vlosfour$, but depend only on $\beta$ and not on its fourth-order counterparts \citep{1990AJ.....99.1548M,2014MNRAS.441.1584R,2017MNRAS.471.4541R}. Thus, $\vsone$ and $\vstwo$ allow us to obtain constraints on $\beta$ via line-of-sight velocities alone, breaking the $\rho-\beta$ degeneracy (e.g. \citealt{1982MNRAS.200..361B,1990AJ.....99.1548M,2017MNRAS.471.4541R}). We use VSPs in our modelling throughout this paper.

We introduce a key improvement in our estimators for $\vsone$ and $\vstwo$ as compared to \citet{2017MNRAS.471.4541R}. In \citet{2017MNRAS.471.4541R}, we assumed that $\vlosfour$ is zero wherever we have no data. This can lead to bias in $\vsone$ and $\vstwo$ if $\vlosfour$ is flat or rising beyond the outermost datapoint, $R_{\rm data}$. To improve on this, we fit a power law to $\vlosfour$ over all radii $R > R_{1/2}$, using this to extrapolate its large $R$ behaviour:

\begin{equation}
\vlosfour = \left\{
\begin{array}{ll}
A \left(\frac{R}{R_{\rm data}}\right)^{-\eta} & R_{\rm data} < R < R_{\rm out} \\ 
A \left(\frac{R_{\rm out}}{R_{\rm data}}\right)^{-\eta} \left(\frac{R}{R_{\rm out}}\right)^{-\kappa}& R > R_{\rm out}
\end{array}
\right .
\end{equation} 
where $A$ and $-2 < \eta < 2$ are fitting parameters and $R_{\rm out}$ sets the outer `edge' of the galaxy. We assume flat priors on $R_{\rm out}$ of $R_{\rm data} < R_{\rm out} < 2 R_{\rm data}$, and on the fall-off of $\vlosfour$ beyond $R_{\rm out}$ of $1 < \kappa < 3$. To determine errors on $\vsone$ and $\vstwo$, we fit the above power law to each of 1000 Monte Carlo draws of the error distribution of $\vlosfour$, as in \citet{2017MNRAS.471.4541R}, marginalising over $R_{\rm out}$ and $\kappa$. In this way, if either $\vsone$ or $\vstwo$ are sensitive to the (unmeasured) large $R$ behaviour of $\vlosfour$, then the errors on these quantities will simply grow. If the data are good enough, however, then the above marginalisation will little affect $\vsone$ and/or $\vstwo$. In tests, we found that the above scheme produces less bias for mocks where $\vlosfour$ is steeply rising at the outermost data point. We will demonstrate its performance on three tidally stripped mocks in \S\ref{sec:mock}.

\GravSphere\ fits the above model to the surface density profile of tracer stars, $\Sigma_*(R)$, their line-of-sight projected velocity dispersion profile $\siglos(R)$ and their VSPs using the \EMCEE\ affine invariant Markov Chain Monte Carlo (MCMC) sampler from \citet{2013PASP..125..306F}. We assume uncorrelated Gaussian errors such that the Likelihood function is given by $\mathcal{L} = \exp(-\chi^2/2)$, where $\chi^2$ includes the contributions from the fits to $\Sigma_*, \siglos$ and the two VSPs.  We use as default 1000 walkers, each generating 5000 models and we throw out the first half of these as a conservative `burn in' criteria. (See \citet{2017MNRAS.471.4541R} for further details of our methodology and priors.)

\subsection{The SIDM model}\label{sec:sidmmodel}

In addition to \GravSphere's default free-form mass model (\S\ref{sec:gravsphere}), we also implement a mass model that describes dark matter halos in a $\Lambda$SIDM cosmology. While more restrictive than our free-form model, this has the advantage that the model parameters correspond to cosmologically interesting quantities like the dark matter halo mass and the SIDM self-interaction cross section. Following \citet{2016MNRAS.461..710D}, \citet{2017MNRAS.470.1542S} and \citet{2017arXiv170501820C}, we consider a velocity-independent SIDM model with an interaction cross section given by:

\begin{equation}
\frac{\sigma}{m} = \frac{\sqrt{\pi}\,\Gamma}{4 \rho_{\rm NFW}(x) \sigma_v(x)}
\label{eqn:sigmam}
\end{equation}
where $\Gamma$ is the SIDM interaction rate, $x$ is a normalisation scale for $\Gamma$ (of which more shortly), $\rho_{\rm NFW}(x)$ is the NFW dark matter density at $x$ before SIDM core formation (equation \ref{eqn:rhoNFW}), and:

\begin{equation}
\sigma_v^2(x) = \frac{G}{\rho_{\rm NFW}}\int_{x}^{\infty} \frac{M_{\rm NFW}(r')\rho_{\rm NFW}(r')}{r'^2}dr'
\label{eqn:sigmavrc}
\end{equation}
is the velocity dispersion of the dark matter halo at $x$ before SIDM core formation. (We assume in this model that the velocity distribution is isotropic.)

We use a \coreNFW\ profile (equation \ref{eqn:rhocNFW}) with $n=1$ to describe the radial density profile of halos in SIDM, as in \citet{2017MNRAS.470.1542S}. However, since we are interested here in satellite galaxies that may have had their outer dark matter density steepened by tidal stripping, we modify the \coreNFW\ profile to account for this, obtaining a new `\coreNFWtides' model:

\begin{equation}
\rho_{\rm cNFWt}(r) = 
\left\{
\begin{array}{ll}
\rho_{\rm cNFW} & r < r_t \\
\rho_{\rm cNFW}(r_t) \left(\frac{r}{r_t}\right)^{-\delta} & r > r_t 
\end{array}
\right.
\label{eqn:coreNFWtides}
\end{equation} 
where $\rho_{\rm cNFW}$ is as in equation \ref{eqn:rhocNFW}, $r_t$ sets the radius at which mass is tidally stripped from the galaxy, and $\delta$ sets the logarithmic density slope beyond $r_t$.

The \coreNFWtides\ model has a number of advantages over previous fitting functions used in the literature. Firstly, it is fully analytic with cumulative mass given by: 

\begin{equation}
M_{\rm cNFWt}(<r) =
\left\{
\begin{array}{ll}
  M_{\rm cNFW}(<r) & r < r_t  \vspace{4mm}\\
M_{\rm cNFW}(r_t) \,\, + & \\
4\pi \rho_{\rm cNFW}(r_t) \frac{r_t^3}{3-\delta}
\left[\left(\frac{r}{r_t}\right)^{3-\delta}-1\right] & r > r_t
\end{array}
\right.
\label{eqn:McNFWt}
\end{equation} 
where $M_{\rm cNFW}$ is as in equation \ref{eqn:coreNFW}. Secondly, it retains the physical meaning of $M_{200}$ and $c_{200}$ in the NFW profile (equation \ref{eqn:rhoNFW}), while introducing two new physically motivated parameters, $r_t$ and $\delta$, to model the effect of tidal stripping beyond $r_t$.

We calibrate the above SIDM model using the \citet{2012MNRAS.423.3740V} pure dark matter cosmological zoom simulations of Milky Way-mass halos in a $\Lambda$SIDM cosmology. Our goal is to ensure that our SIDM model correctly recovers the density profile and scatter of the 15 most massive subhalos in these simulations, since these are the subhalos in which Draco is most likely to reside \citep{2013MNRAS.431L..20Z}. Our free parameters in the calibration are the interaction rate, $\Gamma$, and its normalisation scale, $x$ (see equation \ref{eqn:sigmam}). Typically, $x$ is taken to be the dark matter core size, $x \sim r_c$, for which the interaction rate required to produce a core on the scale of $r_c$ in a Hubble time is of order unity, $\Gamma \sim 1$\,Gyr$^{-1}$ \citep[e.g.][]{2012MNRAS.423.3740V,2016MNRAS.461..710D}. However, in tests we found that, for our \coreNFWtides\ model, using equation \ref{eqn:sigmam} with $x = r_c$ gives a poor fit to the density profiles of subhalos in the \citet{2012MNRAS.423.3740V} SIDM simulations for any choice of constant $\Gamma$. From inspection of equation \ref{eqn:sigmam}, we can understand why this occurs. Notice that, for a constant $\sigma/m$, the density and dispersion at $r_c$ will both fall as the SIDM core forms, lowering $\Gamma$. Since the rate at which this occurs depends on $\sigma/m$, there is no choice of constant $\Gamma$ that can simultaneously fit simulations with low and high $\sigma/m$. We can, however, solve this problem by normalising $\Gamma$ instead at some larger radius $x \gg r_c$ at which the density profile and dispersion change very little after SIDM core formation. We find that $x = 10\,r_c$ with $\Gamma = 0.005$\,Gyr$^{-1}$ gives a good fit to the density profiles of SIDM subhalos in the \citet{2012MNRAS.423.3740V} simulations\footnote{In fact, it is the ratio $\Gamma/\rho_{\rm NFW}(x)$ that is important (see equation \ref{eqn:sigmam}). Since at large $x$, $\rho_{\rm NFW} \propto x^{-3}$, we can find equivalently good fits for $x > 10\,r_c$ and $\Gamma = 0.005\,(10\,r_c/x)^{-3}$\,Gyr$^{-1}$.}. 

\begin{figure}
\begin{center}
\includegraphics[width=0.425\textwidth]{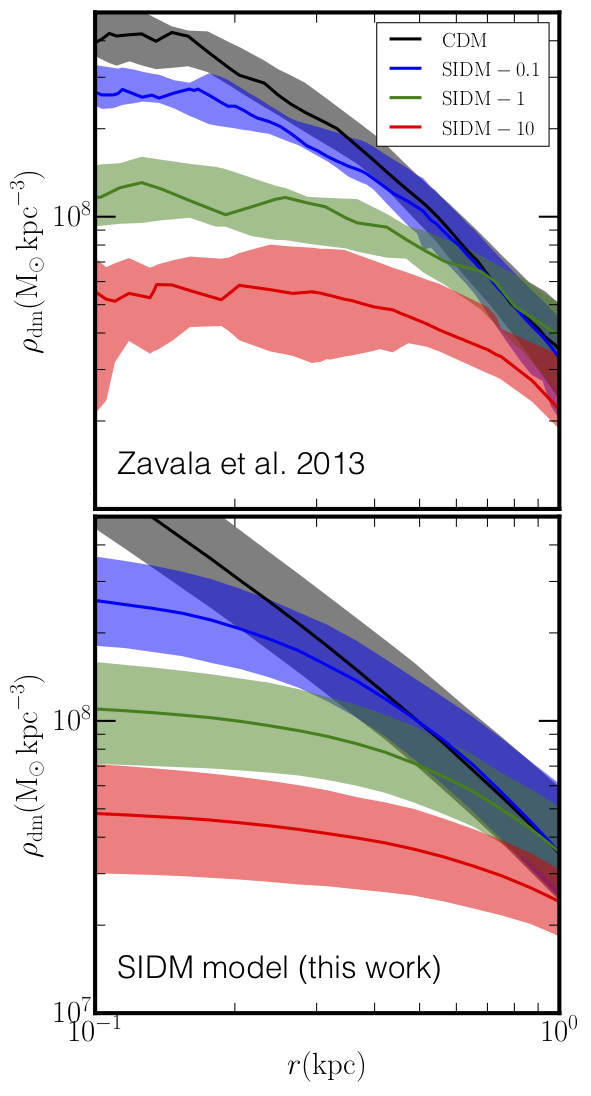}
\vspace{-3mm}
\caption{Calibrating our SIDM model. The panels show the median density profiles (solid lines) and $1\sigma$ scatter (contours) for the 15 most massive subhalos from the \citet{2012MNRAS.423.3740V} and \citet{2013MNRAS.431L..20Z} $\Lambda$SIDM simulations (top), and from our SIDM model (bottom; see text for details). The black lines and contours show results for a $\Lambda$CDM cosmology, while the other coloured lines and bands show results for a $\Lambda$SIDM cosmology with a SIDM cross section of $\sigma/m = 0.1, 1$ and $10$\,cm$^2$/g, as marked. As can be seen, our SIDM model is in good agreement with the median density profile and scatter of subhalos in the \citet{2012MNRAS.423.3740V} simulations.}
\label{fig:SIDM_calibrate}
\end{center}
\end{figure}

To demonstrate that our SIDM model provides a faithful reproduction of subhalos in the \citet{2012MNRAS.423.3740V} simulation, we perform 50 random draws of the 15 most massive subhalos from the subhalo distribution function described in\footnote{The \citet{2012MNRAS.423.3740V} simulations are based on the {{\tt Aquarius}} $\Lambda$CDM simulations described in \citet{2008MNRAS.391.1685S} and have, therefore, similar subhalo statistics \citep{2013MNRAS.431L..20Z}.} \citet{2008MNRAS.391.1685S}. For each halo, we draw its concentration from the $M_{200}-c_{200}$ relation of \citet{2014MNRAS.441.3359D}, multiplied by a factor of 1.4 to account for the increased concentration of subhalos as compared to field halos \citep{2008MNRAS.391.1685S}, and allowing for a scatter of 0.1\,dex. We then use our SIDM model, above, to calculate the radial density profile of each halo for a given choice of $\sigma/m$. Marginalising over all drawn halos, we calculate the median radial density profiles for each $\sigma/m$ and their 68\% confidence intervals. The results are shown in Figure \ref{fig:SIDM_calibrate}. The panels show the median density profiles (solid lines) and $1\sigma$ scatter (contours) for the 15 most massive SIDM subhalos in the \citet{2012MNRAS.423.3740V} SIDM simulations\footnote{This panel was adapted from a Figure in \citet{2013MNRAS.431L..20Z}.} (top) and from our SIDM model (bottom), calculated as described above. The black lines and contours show results for a $\Lambda$CDM cosmology, while the other coloured lines and bands show results for a $\Lambda$SIDM cosmology with a SIDM cross section of $\sigma/m = 0.1, 1$ and $10$\,cm$^2$/g, as marked. As can be seen, our SIDM model is in good agreement with the median density profile and scatter of subhalos in the \citet{2012MNRAS.423.3740V} simulations.

Note that our SIDM model is calibrated on one particular set of SIDM simulations. As such, it is not clear if it can be successfully extrapolated to velocity dependent cross sections, field halos, or halos with a very different mass to those studied here. Furthermore, the model does not include the effect of core collapse. Core collapse can occur for high cross sections ($\sigma/m \simgt 10$\,cm$^2$/g), leading to a steep central density at late times \citep[e.g.][]{2002ApJ...568..475B,2012MNRAS.423.3740V}. We discuss this further in \S\ref{sec:discussion}. Finally, our model does not include the effect of the baryonic potential on the SIDM density profile \citep[e.g.][]{2014PhRvL.113b1302K,2018MNRAS.476L..20R}. This latter is not likely, however, to be a significant effect for Draco whose mass profile is dominated at all radii by dark matter \citep[e.g.][]{2001ApJ...563L.115K}.

When using the \coreNFWtides\ model to describe SIDM halos, we fit the following free parameters: the halo mass and concentration before infall: $M_{200}$ and $c_{200}$; the dark matter core-size parameter $r_c$; the tidal stripping radius $r_t$ and the logarithmic density slope beyond $r_t$, $\delta$. We assume flat priors of $8.75 < \log_{10}(M_{200}/{\rm M}_\odot) < 10.25$; $10 < c_{200} < 22$; $-2 < \log_{10}(r_c/{\rm kpc}) < 0.5$; $0.3 < \log_{10}(r_t / R_{1/2}) < 1$; and $3.5 < \delta < 5$ on these parameters. Our results are not sensitive to these choices. (In particular, extending the upper bound on $c_{200}$ to be $c_{200} < 40$ to account for the higher concentration of subhalos in $\Lambda$CDM (\citealt{2008MNRAS.391.1685S}, and see the discussion above) produces no appreciable effect on our results.)

\section{Testing \GravSpherebf\ on tidally stripped mock data}\label{sec:mock}

\subsection{Three tidally stripped mock Dracos}
\GravSphere\ was extensively tested on mock data in \citet{2017MNRAS.471.4541R}, including on triaxial mocks for which \GravSphere\ (that assumes spherical symmetry) is expected to become biased. However, we did not test \GravSphere\ on tidally stripped mock dwarfs. Such a test is relevant for our paper here since Draco orbits close to the Milky Way (see \S\ref{sec:intro}). We may worry, then, that tidal stripping will induce aspherical distortions and departures from equilibrium that could cause \GravSphere\ to become biased \citep[e.g.][]{2013MNRAS.431.2796K}. To test whether this is an issue for the \GravSphere\ models we present in this paper, we set up three mocks, designed to mimic the Draco dwarf spheroidal galaxy as closely as possible: Mock-Cusp, Mock-Core and Mock-CoreDen, summarised in Table \ref{tab:mock}. 

The Mock-Cusp mock is designed to simulate a Draco with a `pristine' dark matter cusp (see \S\ref{sec:intro}). The Mock-Core model is identical to the Mock-Cusp model, but with a constant density dark matter core of size $\sim R_{1/2}$. This core is consistent with complete core formation in the \citet{2016MNRAS.459.2573R} model, where the core owes to `dark matter heating' due to bursty stellar feedback. It is also consistent with the SIDM model that we described in \S\ref{sec:sidmmodel}, corresponding to a self-interaction cross section of $\sigma/m = 0.22$\,cm$^2/{\rm g}$. However, while the Mock-Core model is cosmologically realistic, its lower central density than the Mock-Cusp model makes it more susceptible to tidal stripping and shocking \citep[e.g.][]{2006MNRAS.366..429R}. For this reason, we include also the Mock-CoreDen model. This is substantially more concentrated than would be expected in either a $\Lambda$CDM or $\Lambda$SIDM cosmology, but may occur in other cosmological models. With the same initial density as the Mock-Cusp model at $\sim 50$\,pc, the Mock-CoreDen model allows us to test \GravSphere\ on a cored mock that has experienced less tidal distortions than the Mock-Core model.

\begin{table}
\begin{center}
\begin{tabular}{l l l c c}
\hline
{\bf Label} & \textbf{\textsc{coreNFW} parameters} & $T_{\rm sim}$ & $\mathbf{r_{tp}}$ \\
 & \vspace{1mm}$[M_{200}({\rm M}_\odot),c_{200},n,r_c({\rm kpc})]$ & [${\rm Gyrs}$] & [{\rm kpc}] \\
\hline
Mock-Cusp \vspace{1mm}& $5 \times 10^9,14,0,-$ & 10 & 1.5 \\
Mock-Core & $5 \times 10^9,14,1,0.315$ & 10 & 1.5 \\
Mock-CoreDen & $5 \times 10^9,35,1,0.315$ & 4.16 & 2.3 \\
\captionsetup{singlelinecheck=false}
\end{tabular}

\end{center}
\vspace{-6mm}
\caption{Mock data initial conditions. From left to right, the columns give the mock data label, the \coreNFW\ model parameters, the total simulation time, and the tidal stripping radius at pericentre, $r_{tp}$ (calculated as in \citet{2006MNRAS.366..429R} using the `prograde' stripping radius; see text for details). The remaining model parameters for the mocks are identical. All three use the same double-Plummer light profile (equation \ref{eqn:doubleplum}), $N_{\rm DM} = 10^7$ dark matter and $N_* = 2 \times 10^7$ star particles, and force softenings $\epsilon_{\rm DM} = 0.009$\,kpc and $\epsilon_* = 0.005$\,kpc, respectively. The mocks were placed on the same orbit around the live Milky Way model from \citet{2008MNRAS.389.1041R}. See \S\ref{sec:mock} for further details.}
\label{tab:mock}
\end{table}

\subsection{The stellar light profile and dark matter distribution}

All three mocks were set up as a spherical galaxy with a double-Plummer sphere of stars:

\begin{equation}
\rho_* = \frac{3 M_*}{8 \pi} \left[\frac{1}{a_1^3}\left(1 + \frac{r^2}{a_1^2}\right)^{-5/2}+\frac{1}{a_2^3}\left(1 + \frac{r^2}{a_2^2}\right)^{-5/2}\right]
\label{eqn:doubleplum}
\end{equation}
with $M_* = 0.29 \times 10^6$\,M$_\odot$, $a_1 = 0.12$\,kpc and $a_2 = 0.23$\,kpc. 

This was embedded in a \coreNFW\ dark matter halo profile (see \S\ref{sec:cuspcore}). For the Mock-Cusp and Mock-Core mocks, we used $M_{200} = 5 \times 10^9$\,M$_\odot$ and $c_{200} = 14$. For the Mock-CoreDen mock, we used $c_{200} = 35$. All three sampled the dark matter with $N_{\rm DM} = 10^7$ particles and the stars with $N_* = 2 \times 10^7$ particles. We used force softenings for the stars and dark matter of $\epsilon_* = 0.005$\,kpc and $\epsilon_{\rm DM} = 0.009$\,kpc, respectively. (We verified that our results are numerically converged at this resolution by comparing them with a similar simulation run with 1/10th of the number of particles. For a discussion of resolution requirements for tidal stripping simulations, we refer the reader to \citealt{2004ApJ...608..663K}, \citealt{2006MNRAS.367..387R} and \citealt{2017arXiv171105276V}.)

For the Mock-Cusp model, we assumed a perfect NFW profile with $n=0$ (a `pristine' cusp). For the Mock-Core and Mock-CoreDen models, we assumed maximal cores, with $n=1$ and $r_c = 0.315$\,kpc, corresponding to a visible core size (see \S\ref{sec:cuspcore}) of $r_{cv} = R_{1/2}$.

\begin{figure}
\begin{center}
\includegraphics[width=0.45\textwidth]{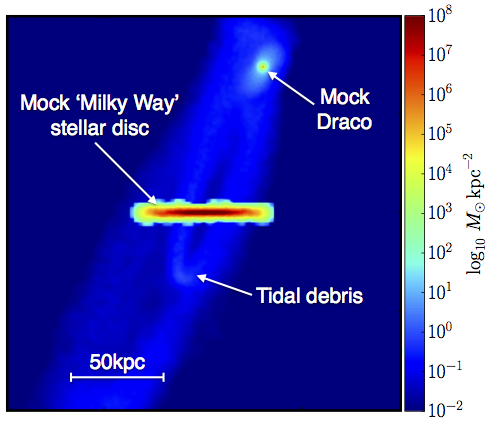}
\caption{A visual impression of the simulations used to produce our mock data. The plot shows the projected density of stars in the Mock-Cusp model. The stellar disc of the host `Milky Way' is seen edge-on, while the dwarf is seen to the top right, as marked. Notice the prominent tidal tails produced as stars are tidally stripped from the mock dwarf by the `Milky Way'.}
\label{fig:mock_snap}
\end{center}
\end{figure}

\begin{figure*}
\begin{center}
\includegraphics[width=0.99\textwidth]{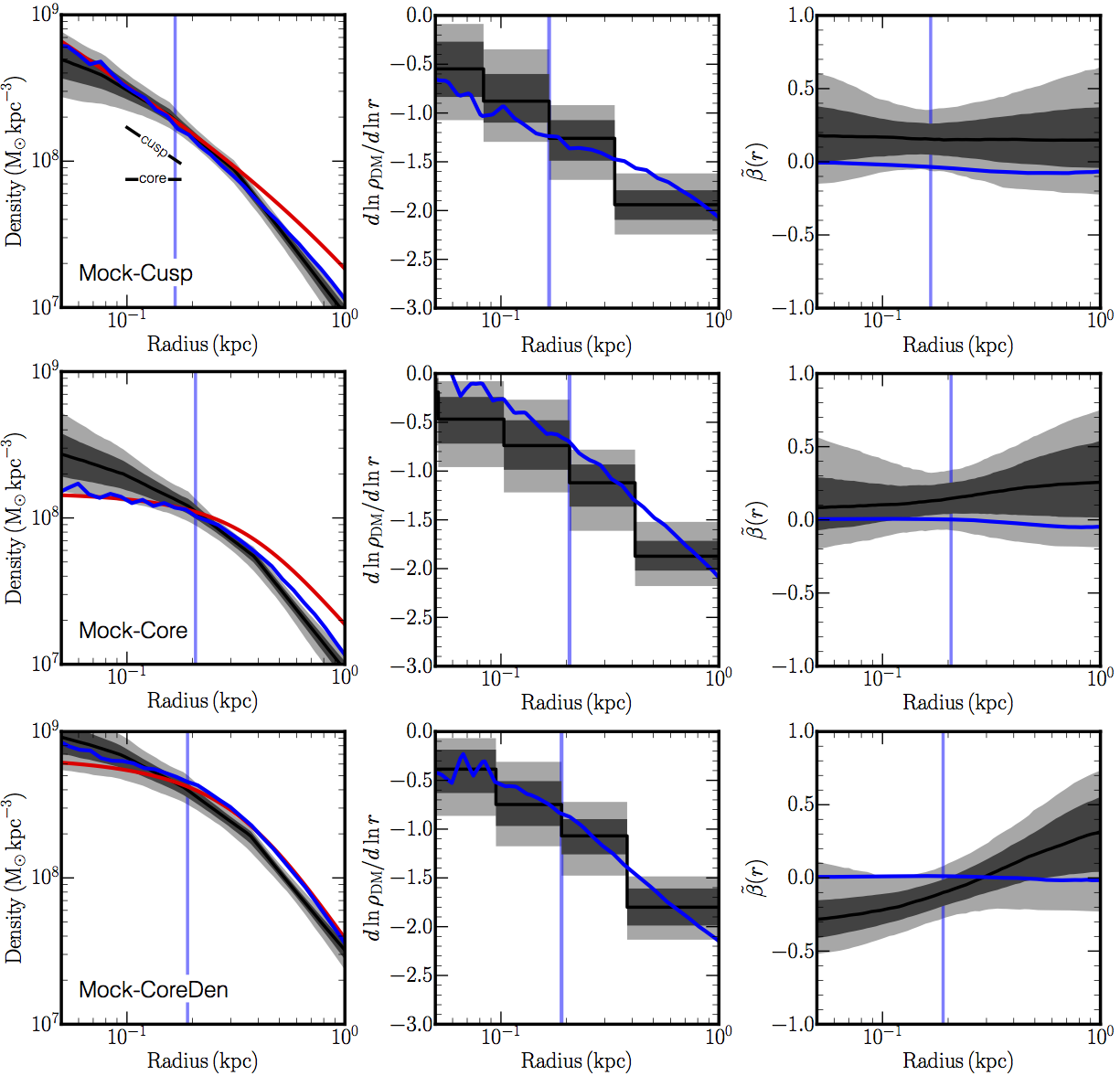}
\caption{\GravSphere\ recovery of the tidally stripped mock dwarfs. From top to bottom, the panels show results for the Mock-Cusp, Mock-Core and Mock-CoreDen mocks, as marked. From left to right, the panels show the spherically averaged dark matter density profile, the logarithmic slope of the density profile ($\gamma_{\rm DM} = d\ln\rho / d\ln r$), and the symmetrised velocity anisotropy profile (equation \ref{eqn:betastar}). The grey contours show the 68\% (dark grey) and 95\% (light grey) confidence intervals of the \GravSphere\ models. The red lines in the left panels show the dark matter density profile of the mocks prior to the action of tides. The blue lines in left panels show the dark matter density profile after tidal stripping and shocking by a `Milky Way'-like galaxy (see \S\ref{sec:mock} for details). This is the `correct answer' that \GravSphere\ should recover. Similarly, the blue lines in the middle and right panels show the correct $\gamma_{\rm DM}$ and velocity anisotropy profiles, respectively. In all panels, the vertical blue lines mark the projected half light radius of the stars, $R_{1/2}$.}
\label{fig:mock}
\end{center}
\end{figure*}

\subsection{The orbit and host Milky Way potential}

The above mock dwarfs were placed on an orbit around a collisionless mock `Milky Way', taken from \citet{2008MNRAS.389.1041R}. This Milky Way model had $N_* = 7.5 \times 10^5$ and $N_{\rm DM} = 2 \times 10^6$, with $\epsilon_* = 0.06$\,kpc and $\epsilon_{\rm DM} = 0.1$\,kpc. The stellar disc had mass and scale length $M_* = 3 \times 10^{10}$\,M$_\odot$ and $r_* = 3$\,kpc, respectively\footnote{The mass of this stellar disc is a factor $\sim 1.5$ lower than that of the Milky Way \citep[e.g.][]{2013ApJ...779..115B}. However, this is not likely to impact our results since neither the mock nor the real Draco comes closer than $\sim 40$\,kpc from the Galactic centre.}, while its dark matter halo was also of NFW form, with $M_{200} = 10^{12}$\,M$_\odot$ and an initial scale length, before growing the disc, of $r_s = 25$\,kpc (after disc growth, the halo contracts yielding a scale length of $r_s \sim 12$\,kpc). The mocks were placed on orbits consistent with Draco's recently measured proper motions \citep{2017arXiv170702593S,2018arXiv180409381G}: ${\bf r}_i = [-4.3,62.3,43.3]$\,kpc and ${\bf v}_i = [57.8,18.2,-172.3]$\,km/s in Galactocentric coordinates. This yields a peri- and apocentre of $r_p = 42$\,kpc and $r_a = 123$\,kpc, respectively, consistent with the orbit calculations in \citet{2017arXiv170702593S}. The Mock-Cusp and Mock-Core models were evolved for 10\,Gyrs using the \pkdgrav\ $N$-body code \citep{2001PhDT........21S}. The Mock-CoreDen model was evolved for 4.16\,Gyrs since longer evolution led to significant numerical relaxation inside $R_{1/2}$. The initial tidal radii of the mocks at pericentre, $r_{tp}$, are reported in Table \ref{tab:mock}. These were calculated as in \citet{2006MNRAS.366..429R}, assuming the `prograde' stripping radius\footnote{\citet{2006MNRAS.366..429R} show that stars moving prograde to the orbit of the satellite around the host galaxy are more easily stripped than stars moving on radial or retrograde orbits (see \citet{1941ApJ....94..385H,1970A&A.....9...24H,1975AJ.....80..290K,2010ApJ...725..353D} and \citet{2016ApJ...819...20G} for earlier and later work on this effect). Over several orbits of the satellite around the host, \citet{2006MNRAS.366..429R} find that stars near the tidal boundary of the satellite have their orbits transformed, leading to a gradual convergence towards the prograde stripping radius.}.

In Figure \ref{fig:mock_snap}, we give a visual impression of the simulations used to produce our mock data. The plot shows the projected density of stars in the Mock-Cusp model. The stellar disc of the host `Milky Way' is seen edge-on, while the dwarf is seen to the top right, as marked. Notice the prominent tidal tails produced as stars are tidally stripped from the mock dwarf by the `Milky Way'. Such tidal stripping occurs in all three mocks, despite their initial tidal radii being substantially larger than their projected half-light radii (see Table \ref{tab:mock}). This occurs as tidal shocks slowly push stars and dark matter over the tidal boundary, gradually whittling the dwarf down \citep[see e.g.][]{2006MNRAS.366..429R}.

\subsection{Binaries, foreground contamination and sampling}

To generate our mock data for \GravSphere, we attempt to mimic as closely as possible the true Draco data. First, we placed each mock Draco at a distance of $D = 82$\,kpc. Next, we discarded all star particles at radii $R < 90$\,arcmins (corresponding to 2.1\,kpc at the distance of Draco). Then, we generated a foreground population of stars with uniform projected density within the $R < 90$\,arcmin field of view. For this foreground population, we assumed ugriz photometry with radial velocities drawn from the Besan\c{c}on model, as implemented in the {\tt Galaxia} code \citep{2003A&A...409..523R,2011ApJ...730....3S}. We then kept only those foreground stars that pass the same isochrone-based filter that was applied to select real Draco targets (see \S\ref{sec:data}). This provides us with a mock `photometric' data set that represents a catalog of RGB candidates. We then sampled spectroscopic quantities for a subset of these mock RGB candidate stars, applying the position-dependent target selection corresponding to the real Draco selection described in \S\ref{sec:data}. We scattered these individual stellar velocities according to errors drawn randomly from the real draco data (typically ${\sim}1$\,km/s). We added binary components to 50\% of the member stars, corresponding to the fraction inferred from multi-epoch data for Draco by Spencer et al. (in preparation), using the distributions of binary orbital elements assumed in that work. Finally, we used the procedure described in \S\ref{sec:data} to obtain membership probabilities for the mock Draco stars. We sampled stars such that the final membership weighted number of mock Draco stars in the photometric and spectroscopic samples were ${\sim}2000$ and ${\sim}500$, respectively, similarly to the real Draco data (see \S\ref{sec:data}). In Appendix \ref{app:mock_bias_core}, we explore the effect of a larger spectroscopic sample size and the influence of binaries and foreground contamination on the Mock-Core mock. There, we show that the binaries and foreground contamination induce some bias in the recovered velocity anisotropy and density profile, but the effect is smaller than our 95\% confidence intervals. A larger spectroscopic sample size leads to tighter constraints on the density profile, as may be expected.

\subsection{Results from applying \GravSphere\ to the mock data}

The \GravSphere\ recovery for all three mocks is shown in Figure \ref{fig:mock}. From left to right, the panels show the spherically averaged dark matter density profile, the logarithmic slope of the density profile ($\gamma_{\rm DM} = d\ln\rho / d\ln r$), and the symmetrised velocity anisotropy profile (equation \ref{eqn:betastar}). The grey contours show the 68\% (dark grey) and 95\% (light grey) confidence intervals of the \GravSphere\ models. The red lines in the left panels show the dark matter density profile of the mocks prior to the action of tides. The blue lines in left panels show the dark matter density profile after tidal stripping and shocking by a `Milky Way'-like galaxy (see \S\ref{sec:mock} for details). This is the `correct answer' that \GravSphere\ should recover. The blue lines in the middle and right panels show, similarly, the correct $\gamma_{\rm DM}$ and velocity anisotropy profiles, respectively. In all panels, the vertical blue lines mark the projected half light radius of the stars, $R_{1/2}$. We show example fits to the data for the Mock-Core and Mock-Cusp mocks in Appendix \ref{app:example_fit}.

For the Mock-Cusp dwarf, \GravSphere\ recovers the input density distribution within its 68\% confidence intervals (see Figure \ref{fig:mock}, left panel, top row). \GravSphere\ correctly detects that this mock dwarf has a high central density of $\rho_{\rm DM}(150\,{\rm pc}) = 2.1_{-0.4}^{+0.5} \times 10^8$\,M$_\odot$\,kpc$^{-3}$ at 95\% confidence, consistent with a $\Lambda$CDM cusp (see \S\ref{sec:cuspcore}), and that its outer density beyond $R_{1/2}$ has been steepened by tidal stripping. The logarithmic slope of the density profile, $\gamma_{\rm DM}(r)$, is recovered within \GravSphere's 68\% confidence intervals (middle panel, top row); \GravSphere\ finds $\gamma_{\rm DM}(150\,{\rm pc}) = -0.89_{-0.25}^{+0.28}$ as compared to the input model, $\gamma_{\rm DM,true}(150\,{\rm pc}) = -1.2$. There is, however, some weak radial bias in the symmetrised velocity anisotropy profile (right panel, top row).

The Mock-Core dwarf is more challenging because of the larger effect of both tidal stripping and shocking. These cause the evolved dark matter density profile to separate from the input model at radii $R \simgt R_{1/2}$ (compare the blue and red lines in Figure \ref{fig:mock}, left panel, middle row). Nonetheless, \GravSphere\ correctly recovers the input model within its 95\% confidence intervals. \GravSphere\ correctly detects that this mock has a low central density of $\rho_{\rm DM}(150\,{\rm pc}) = 1.3_{-0.7}^{+0.6} \times 10^8$\,M$_\odot$\,kpc$^{-3}$ at 95\% confidence, consistent with a small dark matter `core' within $R_{1/2}$ (see \S\ref{sec:cuspcore}). The logarithmic slope of the density profile, $\gamma_{\rm DM}(r)$ (middle panel, middle row) is recovered within \GravSphere's 68\% confidence intervals, but there is a small systematic bias towards cuspier models. \GravSphere\ finds $\gamma_{\rm DM}(150\,{\rm pc}) = -0.72_{-0.26}^{+0.27}$ as compared to the input model, $\gamma_{\rm DM,true}(150\,{\rm pc}) = -0.54$. In tests, we found that this bias is present in 100 random realisations of the Mock-Core mock and so does not owe to an unfortunate random draw. Instead, the bias owes to our choice of priors on $\gamma_{\rm DM}$. We will explore this further in \S\ref{sec:priors}.

Finally, the Mock-CoreDen model presents a challenge not because of tides (it is almost completely immune to tidal effects due to its high density), but because its $\siglos$ and $\vlosfour$ rise steeply to large radii making it more challenging to obtain an unbiased estimate of $\vstwo$ (see \S\ref{sec:method}). For this reason, there is some bias in the recovery of the density profile for this mock (bottom left panel), though the effect is small. The logarithmic slope of the density profile, $\gamma_{\rm DM}$ (Figure \ref{fig:mock}, middle panel, bottom row), is recovered within the 68\% confidence intervals of the \GravSphere\ model chains. \GravSphere\ finds $\gamma_{\rm DM}(150\,{\rm pc}) = -0.75_{-0.22}^{+0.24}$ as compared to the input model, $\gamma_{\rm DM,true}(150\,{\rm pc}) = -0.69$. The symmetrised velocity anisotropy (Figure \ref{fig:mock}, right panel, bottom row) is slightly biased towards tangential models at the centre and radial models at large radii.

\subsection{The effect of our priors on $\gamma_{\rm DM}$}\label{sec:priors}

In this section, we explore the sensitivity of our results to our choice of priors on $\gamma_{\rm DM}$. Our default priors constrain $\gamma_{\rm DM}$ to lie in the range $-3 < \gamma_{\rm DM} < 0$ for each mass bin (see \S\ref{sec:method}). In the absence of sufficiently constraining data, this could cause \GravSphere\ to disfavour cores ($\gamma_{\rm DM} = 0$) because they occupy a smaller hypervolume of the solution space than cusps ($\gamma_{\rm DM} = -1$). To test this, we introduce a rather extreme prior on $\gamma_{\rm DM}$ designed to bias us towards cored models. We assume a flat prior over the range $-3 < \gamma_{\rm DM}' < 2$, and set $\gamma_{\rm DM} = 0$ if $\gamma_{\rm DM}' > 0$ and $\gamma_{\rm DM} = \gamma_{\rm DM}'$ otherwise. In the absence of constraining data, this biases \GravSphere\ towards cores by creating a large region of hypervolume in which $\gamma_{\rm DM} = 0$. Note that we consider this prior to be extreme and use it only to test our sensitivity to priors on $\gamma_{\rm DM}$.

\begin{figure*}
\begin{center}
\includegraphics[width=0.6\textwidth]{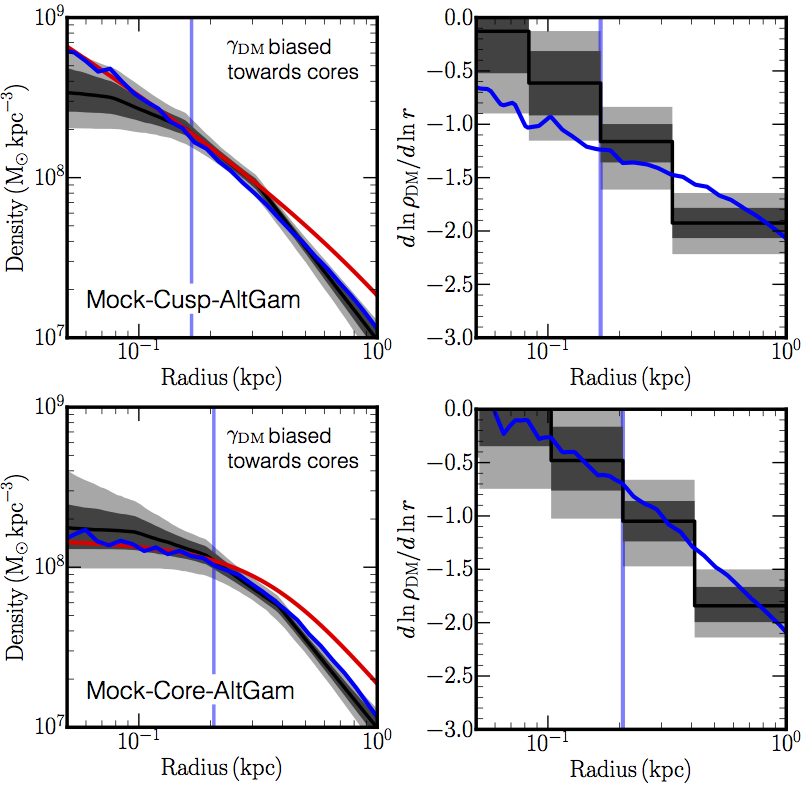}
\caption{Testing the effect of our priors on $\gamma_{\rm DM}$. The plots show the dark matter density profile (left) and logarithmic density slope (right) of \GravSphere\ models for the Mock-Cusp (top) and Mock-Core (bottom) mocks, but using a rather extreme prior on $\gamma_{\rm DM}$ that biases us towards cores (see text for details). The lines and contours are as in Figure \ref{fig:mock}. Notice that our inference of $\gamma_{\rm DM}(r)$ (right panels) is affected by our choice of prior, with the Mock-Cusp mock now being biased towards cored models, while the Mock-Core mock is no longer biased (compare these results with those in the middle panels of Figure \ref{fig:mock}). However, our inference of the amplitude of the inner density at 150\,pc is unaffected by this change in the priors. We obtain $\rho_{\rm DM}(150\,{\rm pc}) = 1.3_{-0.4}^{+0.4} \times 10^8$\,M$_\odot$\,kpc$^{-3}$ at 95\% confidence for the Mock-Core mock and $\rho_{\rm DM}(150\,{\rm pc}) = 2.1_{-0.4}^{+0.5} \times 10^8$\,M$_\odot$\,kpc$^{-3}$ at 95\% confidence for the Mock-Cusp mock, consistent with our default prior estimates.}
\label{fig:mock_gambias}
\end{center}
\end{figure*}

In Figure \ref{fig:mock_gambias}, we rerun the Mock-Cusp (top) and Mock-Core (bottom) mocks using the above modified prior on $\gamma_{\rm DM}$. Notice that our inference of $\gamma_{\rm DM}(r)$ is affected by our choice of prior, with the Mock-Cusp mock now being biased towards cored models (top right panel), while the Mock-Core mock is no longer biased (bottom right panel). However, our inference of the amplitude of the inner density at 150\,pc is unaffected by this change in the priors. We obtain $\rho_{\rm DM}(150\,{\rm pc}) = 1.3_{-0.4}^{+0.4} \times 10^8$\,M$_\odot$\,kpc$^{-3}$ at 95\% confidence for the Mock-Core mock and $\rho_{\rm DM}(150\,{\rm pc}) = 2.1_{-0.4}^{+0.5} \times 10^8$\,M$_\odot$\,kpc$^{-3}$ at 95\% confidence for the Mock-Cusp mock, consistent with our default prior estimates. Furthermore, in all cases -- independently of our choice of prior -- we recover $\rho_{\rm DM}(r)$ and $\gamma_{\rm DM}(r)$ within our 95\% confidence intervals. In Appendix \ref{app:mock_bias_core}, we show that this sensitivity of $\gamma_{\rm DM}$ to our choice of priors diminishes with improved spectroscopic sampling and, therefore, improved constraints on the inner density profile. Finally, note that for 500 stars with spectroscopic velocities the bias on $\gamma_{\rm DM}(R < R_{1/2})$ due to our choice of priors is small, shifting our results by of order the size of our 68\% confidence intervals, even for this rather extreme choice of prior (compare the middle panels in Figure \ref{fig:mock} with the right panels in Figure \ref{fig:mock_gambias}). We will discuss this further when presenting our results for Draco in \S\ref{sec:results}.

\subsection{Testing the recovery of SIDM model parameters using mock data}\label{sec:SIDMtest}

\begin{figure}
\begin{center}
\includegraphics[width=0.45\textwidth]{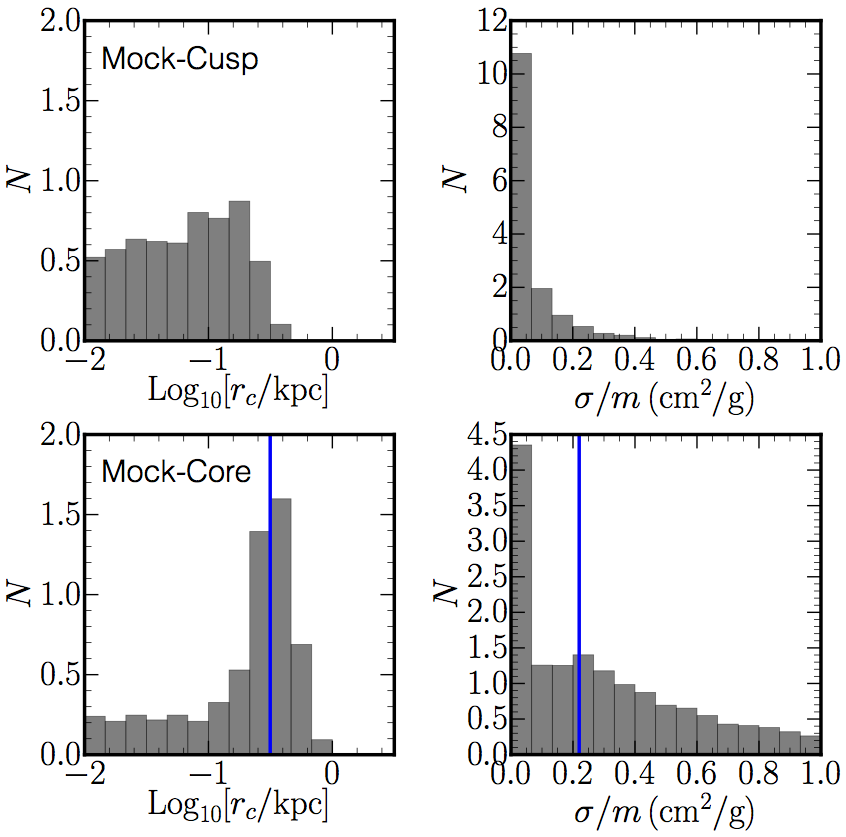}
\caption{Testing the recovery of the SIDM model parameters using the Mock-Cusp (top) and Mock-Core (bottom) mocks. The left panels show the marginalised histograms of the core-size parameter, $r_c$, in the \coreNFWtides\ model (equation \ref{eqn:coreNFWtides}). The right panels show the same for the SIDM self-interaction cross section, $\sigma/m$. For the Mock-Cusp mock (top panels), the correct answer is $r_c = \sigma/m = 0$, while for the Mock-Core model, it is $r_c = 0.315$\,kpc and $\sigma/m = 0.22$\,cm$^2/{\rm g}$, as marked by the vertical blue lines.}
\label{fig:mock_SIDM}
\end{center}
\end{figure}

In this section, we test whether \GravSphere\ is able to correctly recover the SIDM model parameters from our Mock-Core and Mock-Cusp mocks. For this test, we apply \GravSphere\ to the mock data, but using the SIDM mass model described in \S\ref{sec:sidmmodel} rather than \GravSphere's default free-form mass model (\S\ref{sec:gravsphere}). The results are shown in Figure \ref{fig:mock_SIDM}. The left panels show the marginalised histograms of the core size parameter, $r_c$, in the \coreNFWtides\ model (equation \ref{eqn:coreNFWtides}). The right panels show the same for the SIDM self-interaction cross section, $\sigma/m$. For the Mock-Cusp mock (top panels), the correct answer is $r_c = \sigma/m = 0$, while for the Mock-Core model, it is $r_c = 0.315$\,kpc and $\sigma/m = 0.22$\,cm$^2/{\rm g}$, as marked by the vertical blue lines. Notice than in both cases, $r_c$ and $\sigma/m$ are well-recovered. For the Mock-Cusp mock, this translates into upper bounds on both parameters, since a small core inside $\sim 0.5 R_{1/2}$ is still permitted within the uncertainties. For the Mock-Core mock, \GravSphere\ well-recovers $r_c$, though there is a weak tail to low $r_c$ cuspy models. This translates into a second peak at low $\sigma/m$ (bottom right panel of Figure \ref{fig:mock_SIDM}). 

\section{Data}\label{sec:data}

As \GravSphere\ fits both surface density and projected velocity dispersion profiles, we require both photometric and kinematic data that sample Draco's stellar population.  For the photometric data we use the Pan-STARRS DR1 catalog \citep{flewelling16}, initially selecting point-like sources\footnote{We select point source objects for which the difference between PSF and Kron magnitudes in the $r$ band is $r_{\rm PSF}-r_{\rm kron}<0.05$ (see \citealt{farrow14} for a discussion of Pan-STARRS star-galaxy separation).} within $1.5^{\circ}$ of Draco's nominal center at $\alpha_{\rm J2000}=\text{17:20:14.4}$, $\delta_{\rm J2000}=\text{+57:54:54}$ \citep{martin08}.   From these point sources we obtain a sample of candidate red giant branch (RGB) stars within Draco by selecting only sources that are brighter than $i\leq 21$ mag and deviate in colour-magnitude ($g-r$, $i$) space by less than $\epsilon$ magnitudes from an old (${\rm age} = 12$\,Gyr), metal-poor (${\rm [Fe/H]} = - 2.5$) model isochrone \citep{dotter08} that we shift by distance modulus $m-M=19.6$, corresponding to Draco's distance of $D \sim 76$ kpc \citep{2012AJ....144....4M}.  For this work, we adopt $\epsilon=\sqrt{0.04+\sigma^2_i+\sigma^2_{g-r}}$, where $\sigma_i$ and $\sigma_{g-r}$ are the Pan-STARRS uncertainties in magnitude and colour, respectively.  This procedure yields a sample of 15,891 RGB candidates with uniform selection out to radius $R\leq 1.5^{\circ}$.

For the stellar-kinematic data we adopt the spectroscopic sample published by \citet{walker15}, which consists of line-of-sight velocities, effective temperatures, surface gravities and metallicities measured for 1,565 RGB and horizontal branch candidates within $1.5^{\circ}$ of Draco's center.  Applying hard cuts to separate members from foreground contamination according to each of these observables, \citet{walker15} estimate that this sample contains $\sim 500$ probable members of Draco.  

In order to achieve a more quantitative separation between Draco members and contamination from the Galactic foreground, we fit an initial, chemodynamical mixture model similar to the one described in detail by \citet{caldwell17} for their analysis of the dwarf galaxy Crater 2.  That is, we fit simultaneously for: 1) the position distribution of RGB candidates in the photometric sample; and 2) the joint distribution of velocities and metallicities of RGB candidates in the spectroscopic sample. Following \citet{caldwell17}, this initial fit assumes that: 1) the positions of Draco members follow a (single-component) Plummer profile, with projected stellar density $\Sigma_{\rm Dra}(R)=L(\pi a^2)^{-1}(1+R^2/a^2)^{-2}$, where $L$ and $a$ are Draco's total luminosity and projected half-light radius, respectively; 2) the velocities, $V$, and metallicities, $Z$, of Draco members follow independent normal distributions: $P_{\rm Dra}(V,Z)=\mathcal{N}(\overline{V},\sigma_V^2+\delta_V^2)\mathcal{N}(\overline{Z},\sigma_Z^2+\delta_Z^2)$, where $\overline{V}$ and $\overline{Z}$ are mean velocity and mean metallicity, $\sigma_Z$ and $\sigma_Z$ are intrinsic velocity and metallicity dispersions, and $\delta_V$, $\delta_Z$ are observational errors; 3) non-members in the Galactic foreground follow a uniform spatial distribution, $\Sigma_{\rm MW} = {\rm constant}$, with velocity and metallicity distributions estimated empirically by smoothing the data with a Gaussian kernel, denoting these estimates $\hat{P}_{\rm MW}(V)$ and $\hat{P}_{\rm MW}(Z)$.  Our initial model is simpler than that of Caldwell et al. (2017), however, in that we assume that any velocity and/or metallicity gradients are negligible. After fitting this model, we evaluate for every star a probability of Draco membership\footnote{The majority of stars in the photometric sample lack spectroscopic velocity and metallicity measurements; for these stars we evaluate membership probability using a simplified model wherein the probabilities of spectroscopically-observed quantities are set equal to unity and the membership probability depends solely on position.}, $P_{\rm mem}(R,V,Z)=M/(M+N)$, where $M\equiv \Sigma_{\rm Dra}(R)P_{\rm Dra}(V,Z)$ and $N\equiv\Sigma_{\rm MW}(R)\hat{P}_{\rm MW}(V)\hat{P}_{\rm MW}(Z)$. Summing these probabilities, we estimate that the photometric sample of RGB candidates contains $N_{\rm mem,phot}=\sum_{i=1}^{N_{\rm phot}}P_{\rm mem,phot,i}=2,500\pm 56$ members of Draco, while the spectroscopic one contains $N_{\rm mem,spec}=\sum_{i=1}^{N_{\rm spec}}P_{\rm mem,spec,i}=504\pm 1$ members.  

We construct empirical surface density and projected velocity dispersion profiles for Draco by dividing the photometric and spectroscopic data sets, respectively, according to projected radius into annular bins. Each of these bins contains an equal membership-probability-weighted number of stars. We adopt $N_{\rm phot}=N_{\rm kin}=15$ for both the surface density profile and the velocity dispersion profile. Figure \ref{fig:example_fit} displays the stellar surface density and projected velocity dispersion profiles that we obtain for Draco.

This construction of the binned profiles is imperfect for several reasons.  First, the profiles reflect only the median posterior probability of membership for each star, and thus do not propagate variance in those membership probabilities.  Second, the membership probabilities are derived from a model that incorporates simplifying assumptions---e.g., that the stellar positions follow a single-component Plummer profile and that the velocities follow a single Gaussian distribution---that are generally inconsistent with the one that \GravSphere\ subsequently fits. A fully consistent treatment would require allowing for position-dependent and/or non-Gaussian velocity distributions and building a background model into our \GravSphere\ analysis, a task that we reserve for future work.  For now, we have confirmed that our results for Draco are qualitatively unchanged if we use for the initial fit a more sophisticated model that is based on the spherical Jeans equation, explicitly includes a dark matter halo and thereby allows the stellar velocity dispersion (and resulting dependence of membership probability on velocity) to vary with radius. For details of this model, see Section 4.5 of \citet{caldwell17}. 

Finally, \citet{2013ApJ...763...91J} obtained a measurement of the inner dispersion profile of Draco at $\sim 5$\,pc from its centre using Virus-P spectrograph velocity measurements for 17 stars, of which 12 were found to be members of Draco. We experimented with including also these data, however they led to no noticeable change in our favoured distribution of models for Draco. As such, we present here results only using the \citet{2015MNRAS.448.2717W} data that are selected and reduced in a fully consistent manner, with a consistent membership criteria.

\section{Results for Draco}\label{sec:results}

\subsection{The dark matter density profile}\label{sec:dmprofile}

\begin{figure*}
\begin{center}
\includegraphics[width=0.99\textwidth]{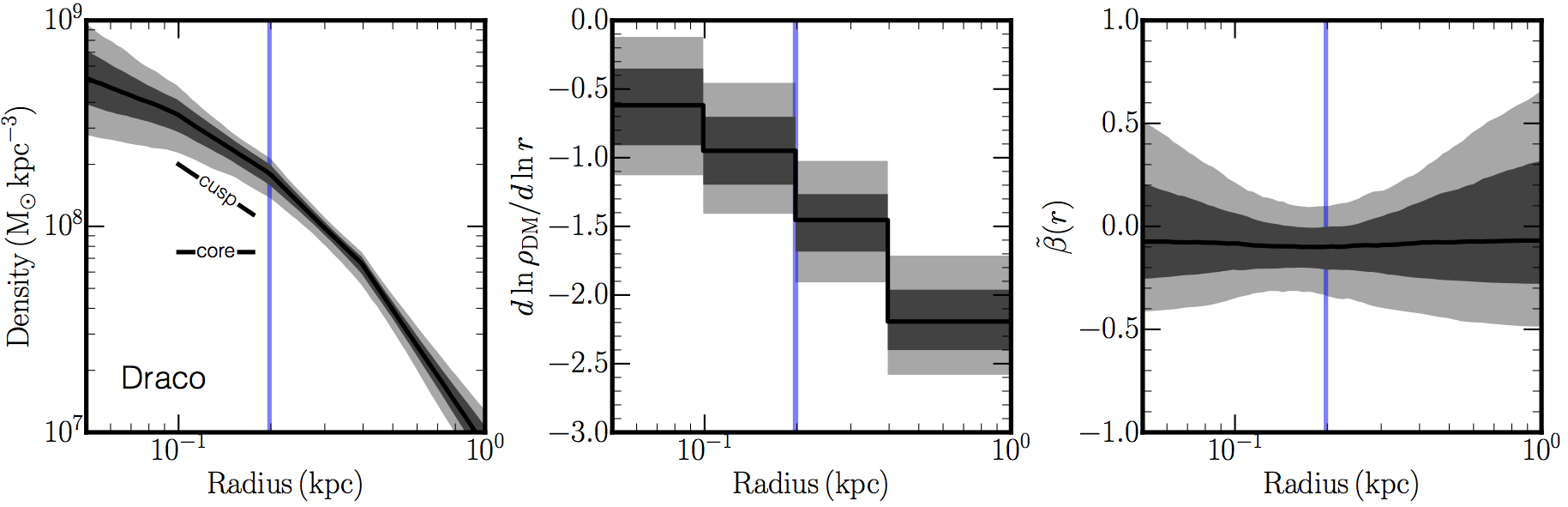}
\caption{As Figure \ref{fig:mock}, but for the real Draco data.}
\label{fig:draco}
\end{center}
\end{figure*}

In this section, we apply \GravSphere\ to the real Draco data from \citet{2015MNRAS.448.2717W} (see \S\ref{sec:data}). The results are shown in Figure \ref{fig:draco}, where the lines and panels are as in Figure \ref{fig:mock}. (We show the \GravSphere\ fits to the projected velocity dispersion, photometric light profile and VSPs in Appendix \ref{app:example_fit}, Figure \ref{app:example_fit}.)

Firstly, notice that -- as for our mock data -- we obtain strong constraints on $\betastar$ only near the projected half stellar mass radius, $R_{1/2}$ (vertical blue line). The \GravSphere\ models for Draco are consistent with velocity isotropy at all radii, similarly to our mock data.

The key result for this paper is the spherically averaged dark matter density profile for Draco (Figure \ref{fig:draco}, left panel). Marked on this panel are a power law cusp and a core. As can be seen, our \GravSphere\ models for Draco are more similar to the cusped model, with a large central density, $\rho_{\rm DM}(150\,{\rm pc}) = 2.4_{-0.6}^{+0.5} \times 10^8\,{\rm M}_\odot \,{\rm kpc}^{-3}$ and logarithmic slope $\gamma_{\rm DM}(150\,{\rm pc}) = -0.95_{-0.46}^{+0.50}$ at 95\% confidence. In Appendix \ref{app:draco_robust}, we show that this high central density is robust to modelling Draco without $\vsone$ or $\vstwo$ (equations \ref{eqn:vs1} and \ref{eqn:vs2}), and to changing the priors on $\gamma_{\rm DM}$. Switching to a rather extreme prior that biases us towards cored models (see \S\ref{sec:priors}), we find $\rho_{\rm DM}(150\,{\rm pc}) = 2.1_{-0.6}^{+0.5} \times 10^8\,{\rm M}_\odot \,{\rm kpc}^{-3}$, with a logarithmic slope of $\gamma_{\rm DM}(150\,{\rm pc}) = -0.7_{-0.52}^{+0.52}$ at 95\% confidence that still favours a cusp.

Our \GravSphere\ models for Draco are in good agreement with pure dark matter structure formation simulations in $\Lambda$CDM (e.g. \citealt{1991ApJ...378..496D,1996ApJ...462..563N}; and \S\ref{sec:intro}). We consider, next, what such a steep cusp in Draco implies for self-interacting dark matter models.

\subsection{A new constraint on the Self-Interacting Dark Matter (SIDM) cross section}\label{sec:sidm}

\begin{figure}
\begin{center}
\includegraphics[width=0.45\textwidth]{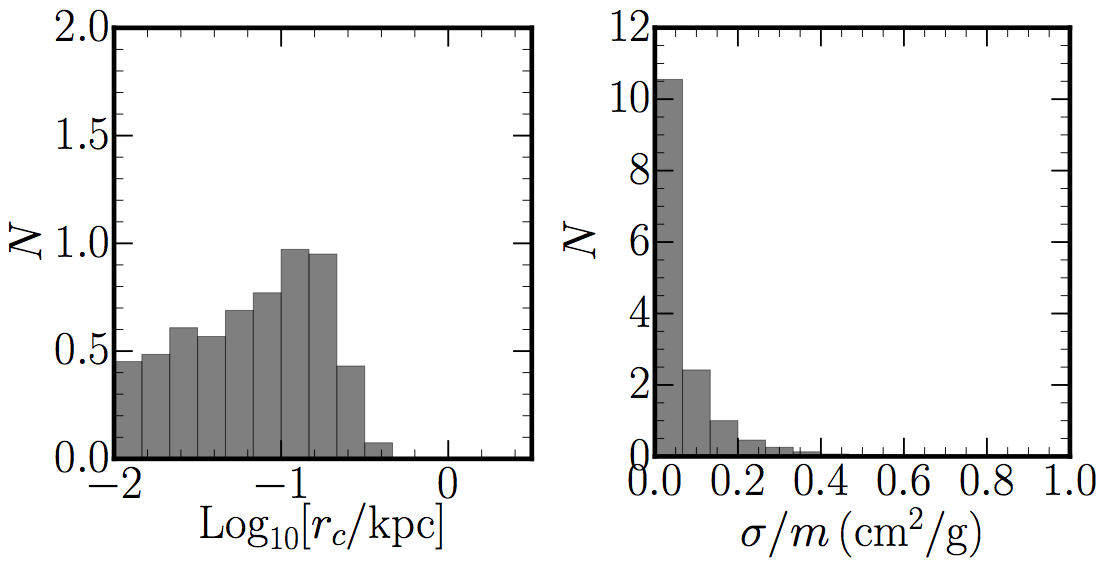}
\caption{A new constraint on the SIDM self-interaction cross section. The left panel shows the marginalised histogram of dark matter core-size parameters, $r_c$, from fitting the SIDM \coreNFWtides\ model to the Draco data (equation \ref{eqn:coreNFWtides}). The right panel shows the corresponding histogram of SIDM self-interaction cross sections $\sigma/m$ for this same model. The high central density that we infer for Draco disfavours models with $\sigma/m > 0.57$\,cm$^2$/g at 99\% confidence.}
\label{fig:DracoSIDM}
\end{center}
\end{figure}

In this section, we fit the SIDM model described in \S\ref{sec:sidmmodel} to the data for Draco to place a new upper bound on the SIDM cross section. The results are shown in Figure \ref{fig:DracoSIDM}. The left panel shows the marginalised histogram of dark matter core-size parameters, $r_c$ (equation \ref{eqn:coreNFWtides}); the right panel shows the corresponding histogram of SIDM self-interaction cross sections, $\sigma/m$, for this same model. Notice that models with large dark matter cores are disfavoured. We find, subject to our choice of SIDM model and prior, $r_c < 0.36$\,kpc at 99\% confidence, corresponding to a visible core size of $r_{cv} < 0.21$\,kpc at 99\% confidence (see \S\ref{sec:mock}). Thus, consistent with our free-form models for Draco, our SIDM models imply that, if Draco has a dark matter core, it is likely smaller than $R_{1/2}$. This upper bound on $r_c$ corresponds to a new constraint on the SIDM self-interaction cross section of $\sigma/m < 0.32$\,cm$^2$/g at 95\% confidence and $\sigma/m < 0.57$\,cm$^2$/g at 99\% confidence. (Recall that the mapping between $r_c$ and $\sigma/m$ depends also on the halo mass and concentration.)

For the other \coreNFWtides\ parameters, we obtain a constraint on the pre-infall halo mass of Draco of $M_{200} = 2.6_{-0.8}^{+1.1} \times 10^9\,{\rm M}_\odot$ at 68\% confidence, consistent with our mock Draco models (\S\ref{sec:mock}). We do not obtain strong constraints on $c_{200}$, $\delta$ nor $r_t$ (see equation \ref{eqn:coreNFWtides} for a definition of these).

\section{Discussion}\label{sec:discussion}

\subsection{Model caveats}

There are three main caveats to our result that Draco has a central dark matter cusp. Firstly, \GravSphere, while being largely assumption-free, still assumes spherical symmetry and dynamic equilibrium. In \citet{2017MNRAS.471.4541R}, we tested \GravSphere\ on triaxial mock data that had triaxiality of the magnitude expected in $\Lambda$CDM. We found that the systematic error that this induces is typically smaller than the 95\% confidence intervals of the \GravSphere\ models. Furthermore, when the systematic bias did become significant, the model fit was poor. We see no evidence of this behaviour for Draco (Figure \ref{fig:example_fit}). This is consistent with other work in the literature that has found that spherical models can successfully recover the radial density profile of triaxial mock data \citep{2013MNRAS.433L..54L,2017arXiv170706303G,2017arXiv170809425K}. In \S\ref{sec:mock}, we showed further that tidally stripped stars in Draco are also unlikely to influence our result. This is in tension with previous findings by \citet{2013MNRAS.431.2796K} who report a significant bias when applying spherical equilibrium models to tidally stripped mocks. This difference could owe to the fact that \citet{2013MNRAS.431.2796K} test simple Jeans mass estimators that are known to be more biased than fully self-consistent dynamical models \citep[e.g.][]{2017MNRAS.469.2335C}, or it could owe to their tidally stripped mocks being much further from equilibrium than the Mock-Cusp, Mock-Core and Mock-CoreDen models that we consider here. It is beyond the scope of this present work to explore this in more detail. 

The second potential caveat to our results is in our choice of data selection and binning. To test the importance of this, we ran a large suite of \GravSphere\ models for Draco varying the data binning and membership selection criteria (see \S\ref{sec:data}). In all cases, we found a central cusp, consistent with that in Figure \ref{fig:draco} (left panel). However, it could be that our assumption of a Gaussian velocity distribution function when calculating the membership probability could bias our results, particularly at large radii where contamination is more problematic (see \S\ref{sec:data}). We will explore this further in future work.

Finally, in \S\ref{sec:mock} we found that with only 500 stars with spectroscopic data, \GravSphere\ was able to distinguish $\rho_{\rm DM}(150\,{\rm pc})$ for our Mock-Cusp and Mock-Core mocks at 95\% confidence. However, the logarithmic slope of the density profile, $\gamma_{\rm DM}(150\,{\rm pc})$, depended on our choice of prior. Using a more conservative prior on $\gamma_{\rm DM}$ that is biased towards cores, we found $\gamma_{\rm DM} < -0.2$ at 95\% confidence, providing only weak evidence for a formal cusp. Increasing the spectroscopic sampling for Draco to $1000-2000$ stars would reduce our sensitivity to the priors on $\gamma_{\rm DM}$ and improve our constraints (see Appendix \ref{app:mock_bias_core}).

\subsection{Comparison with previous work}\label{sec:compare}

Draco has long been known to be one of the densest of the Milky Way dwarfs \citep[e.g.][]{2001ApJ...563L.115K}. For this reason, it consistently features high on the list of targets for dark matter annihilation and decay searches \citep[e.g.][]{2011MNRAS.418.1526C,2015MNRAS.446.3002B,2016PhRvD..93j3512E}. However, only one study to date has reported being able to constrain the central logarithmic slope of Draco's dark matter profile.  \citet{2013ApJ...763...91J} used a non-parametric Schwarzschild method applied to stellar kinematic data near the centre of Draco, obtained with the Virus-P spectrograph. They found a central logarithmic cusp slope of $\gamma_{\rm DM} = -1.0 \pm 0.2$ over the range $20 < r < 700$\,pc. This agrees well with our \GravSphere\ models for Draco. More recently, \citet{2017arXiv171103502V} fit SIDM models to Draco, finding that Draco favours a low $\sigma/m \simlt 0.5\,{\rm cm}^2/{\rm g}$, consistent with our findings here. This latter study is particularly interesting. They fit all of the Milky Way classical dwarfs with an SIDM model, finding a range of central densities that translates into a broad range of favoured $\sigma/m$. We will discuss this further in a companion paper where we apply \GravSphere\ to all of the Milky Way classical dwarfs.

\subsection{A small dark matter core in Draco}\label{sec:smallcore}

We have shown that our \GravSphere\ models for Draco favour a dark matter cusp over the range $100 < R/{\rm pc} < R_{1/2}$. However, this still leaves room for a $\simlt 100$\,pc dark matter core within our \GravSphere\ model uncertainties (Figure \ref{fig:draco}, left panel). This is interesting for two reasons. Firstly, our SIDM model constraints are based on a velocity-independent SIDM model fit to a particular set of SIDM simulations (\S\ref{sec:sidmmodel}). As has been pointed out by several authors, SIDM can have a rather rich and complex dynamics due to, for example, late-time core collapse and tidal effects \citep[e.g.][]{2002ApJ...568..475B,2012MNRAS.423.3740V}. It could be that more detailed SIDM models have smaller central cores that are able to match the data for Draco with larger cross sections than we report here. Secondly, two studies have recently used the survival and properties of dense star clusters in the `ultra-faint' dwarfs Eridanus II \citep{2017ApJ...844...64A,2017arXiv170501820C} and Andromeda XXV \citep{2017ApJ...844...64A} to argue for the presence of dark matter cores. This raises an important question: are the claimed dark matter cores in Eridanus II and Andromeda XXV at odds with our findings here for Draco?

Firstly, note that \citet{2017arXiv170501820C} show that Eridanus II's dark matter core has a size $> 45$\,pc and a density in the range $6 \times 10^7 - 2.5 \times 10^8$\,M$_\odot$\,kpc$^{-3}$. Draco could host a $\simlt 100$\,pc-size dark matter core at the upper end of this range (see Figure \ref{fig:draco}, left panel). Such a core could result from a modification to dark matter (e.g. SIDM with a low self-interaction cross section). However, it seems unlikely that this same model could then be responsible also for the much larger dark matter core reported in Fornax \citep[e.g.][]{GoerdtEtAl2006,2011ApJ...742...20W,2012MNRAS.426..601C,2018arXiv180202606P} and the similarly large cores reported in nearby gas-rich isolated dwarf irregulars \citep[e.g.][]{1994Natur.370..629M,1994ApJ...427L...1F,2017MNRAS.467.2019R}. By contrast, models in which dark matter is heated by bursty stellar feedback could naturally account for such a diversity of central dark matter densities, at least in principle. Recall that whether or not a dark matter core can form from such `heating' depends on: (i) the pre-infall dark matter halo mass, $M_{200}$; (ii) the halo concentration parameter, $c_{200}$; (iii) the total stellar mass, $M_*$; and (iv) the size of the dark matter core \citep[e.g.][]{2014MNRAS.437..415D,2016MNRAS.459.2573R,2017arXiv170501820C,2018MNRAS.479.1514B}. (Smaller cores require less energy to form and form more rapidly.) Dark matter cores are also easier to form at high redshift when the star formation rates are high and the halo masses are smaller \citep{2014ApJ...789L..17M}. This suggests that {\it small} ($\simlt 100$\,pc) dark matter cores may indeed form very early in the Universe even in ultra-faint dwarfs. This effect could be ubiquitous, or it could be stochastic, depending on the merger history, spin and/or concentration of any given dwarf \citep[e.g.][]{2015MNRAS.449L..90L}. We will return to this issue in more detail in a forthcoming paper where we present \GravSphere\ models for all of the Milky Way classical dwarfs.

\section{Conclusions}\label{sec:conclusions}

We have used a new mass modelling method, \GravSphere, to measure the central dark matter density profile of the Draco dwarf spheroidal galaxy. Our key findings are as follows:

\begin{itemize}

\item Using mock data with sampling, binary star population, and foreground contamination similar to that for the real Draco dwarf, we showed that \GravSphere\ is able to successfully recover the dark matter density profile of a tidally stripped Draco-like dwarf within its 95\% confidence intervals (Figure \ref{fig:mock}). However, while we were able to distinguish the amplitude of the central density, $\rho_{\rm DM}(150\,{\rm pc})$, of our cored and cusped mocks at 95\% confidence, the logarithmic slope of the density profile, $\gamma_{\rm DM}(150\,{\rm pc})$, depended on our choice of priors. This sensitivity to the prior diminishes with improved spectroscopic sampling (Appendix \ref{app:mock_bias_core}).

\item We then applied \GravSphere\ to the real Draco data. We inferred a high central density of $\rho_{\rm DM}(150\,{\rm pc}) =  2.4_{-0.6}^{+0.5} \times 10^8\,{\rm M}_\odot \,{\rm kpc}^{-3}$ with logarithmic slope $\gamma_{\rm DM}(150\,{\rm pc}) = -0.95_{-0.46}^{+0.50}$ at 95\% confidence, consistent with expectations from pure-dark matter structure formation simulations in $\Lambda$CDM (Figure \ref{fig:draco}). We tested the robustness of this result, showing that even using a rather extreme prior on $\gamma_{\rm DM}$ that biases us towards cored models, our \GravSphere\ models still favour a logarithmic slope of $\gamma_{\rm DM}(150\,{\rm pc}) = -0.70_{-0.52}^{+0.52}$ at 95\% confidence, steeper than that of a uniform-density core. Dark matter models with a high central density and a shallow inner slope are, however, still permitted.

\item At smaller radii, $R < 0.5 R_{1/2}$, our \GravSphere\ model constraints are poorer, consistent with both a dark matter cusp and a core within our 95\% confidence intervals.

\item We fit a velocity independent SIDM model to the Draco data, obtaining -- subject to our choice of SIDM model and prior -- a new upper bound on the dark matter self-interaction cross section of $\sigma/m < 0.32$\,cm$^2$/g at 95\% confidence and $\sigma/m < 0.57$\,cm$^2$/g at 99\% confidence. This illustrates how Draco's high central density, in combination with constraints on the density profile at larger radii, can be used to constrain interesting dark matter models. We will consider in future work whether such a high density can be consistent with other modifications to dark matter.

\item Finally, our SIDM model fit also provided a constraint on the pre-infall dark matter halo mass of Draco. We found $M_{200} = 2.6_{-0.7}^{+1.1} \times 10^9\,{\rm M}_\odot$ at 68\% confidence.
\end{itemize}

\section{Acknowledgments}
We would like to thank Manoj Kaplinghat, Giuseppina Battaglia, Aurel Schneider, Richard Massey, Andrew Robertson and the anonymous referee for helpful feedback on an early draft of this paper. JIR would like to thank the KITP in Santa Barbara and the organisers of the ``The Small-Scale Structure of Cold(?) Dark Matter'' programme. This paper benefited from helpful discussions that were had during that meeting. JIR would like to acknowledge support from SNF grant PP00P2\_128540/1, STFC consolidated grant ST/M000990/1 and the MERAC foundation. This research was supported in part by the National Science Foundation under Grant No. NSF PHY-1748958. M.G.W. is supported by National Science Foundation grant AST-1412999. Figure \ref{fig:mock_snap} and the mock data analysis made use of the excellent {{\tt PyNbody}} software package\footnote{\url{https://github.com/pynbody/pynbody}.} \citep{pynbody}.

The Pan-STARRS1 Surveys (PS1) and the PS1 public science archive have been made possible through contributions by the Institute for Astronomy, the University of Hawaii, the Pan-STARRS Project Office, the Max-Planck Society and its participating institutes, the Max Planck Institute for Astronomy, Heidelberg and the Max Planck Institute for Extraterrestrial Physics, Garching, The Johns Hopkins University, Durham University, the University of Edinburgh, the Queen's University Belfast, the Harvard-Smithsonian Center for Astrophysics, the Las Cumbres Observatory Global Telescope Network Incorporated, the National Central University of Taiwan, the Space Telescope Science Institute, the National Aeronautics and Space Administration under Grant No. NNX08AR22G issued through the Planetary Science Division of the NASA Science Mission Directorate, the National Science Foundation Grant No. AST-1238877, the University of Maryland, Eotvos Lorand University (ELTE), the Los Alamos National Laboratory, and the Gordon and Betty Moore Foundation.

\appendix
\section{Example \GravSpherebf\ model fits}\label{app:example_fit}

\begin{figure*}
\begin{center}
\includegraphics[width=\textwidth]{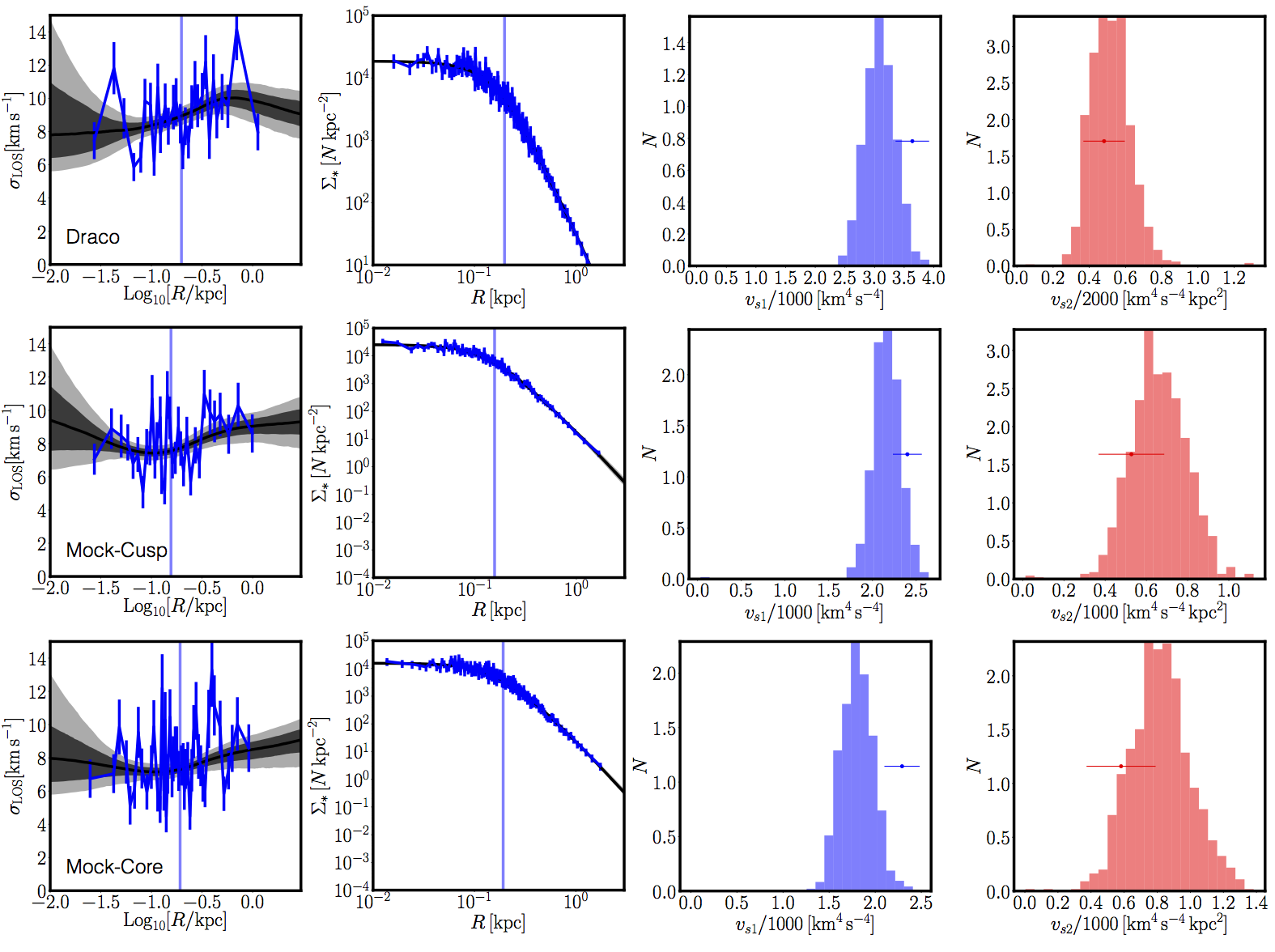}
\vspace{-4mm}
\caption{Example \GravSphere\ model fits for Draco (top), the Mock-Cusp mock (middle) and the Mock-Core mock (bottom). The panels show, from left to right: the projected stellar velocity dispersion profile; the surface brightness profile of the stars; and the `Virial Shape Parameters', $v_{s1}$ (equation \ref{eqn:vs1}) and $v_{s2}$ (equation \ref{eqn:vs2}). The grey contours show the 68\% (dark) and 95\% (light) confidence intervals of the \GravSphere\ model chains. The data points mark the input data used for the model fits. Notice that the errors on $v_{s1}$ and $v_{s2}$ are comparable for the real Draco data and the mock, despite the mock having more kinematic tracers. This occurs because the mock data have a steeply rising $\vlosfour$ to large $R$, unlike the true Draco data (see the text and \S\ref{sec:method} for further details).}
\label{fig:example_fit}
\end{center}
\end{figure*}

In this Appendix, we show example \GravSphere\ model fits for the real Draco data and the Mock-Cusp and Mock-Core mocks. (The fit for the Mock-CoreDen mock is comparably good and so we omit this for brevity.) The results are shown in Figure \ref{fig:example_fit}, where the panels show from left to right: the projected stellar velocity dispersion profile; the surface brightness profile of the stars; and the `Virial Shape Parameters', $v_{s1}$ (equation \ref{eqn:vs1}) and $v_{s2}$ (equation \ref{eqn:vs2}). The grey contours show the 68\% (dark) and 95\% (light) confidence intervals of the \GravSphere\ model chains. The data points mark the input data used for the model fits.

For both the mock data and the real Draco data, the \GravSphere\ models recover the input data within the error bars. In \citet{2017MNRAS.471.4541R} we found, using mock data, that when unmodelled triaxiality caused significant bias in the models, this manifested also in a poor fit to either $v_{s1}$ or $v_{s2}$. This does not appear to be the case for Draco.

\section{Testing the effect of binary stars, foreground contamination and spectroscopic sample size}\label{app:mock_bias_core}

\begin{figure*}
\begin{center}
\includegraphics[width=0.99\textwidth]{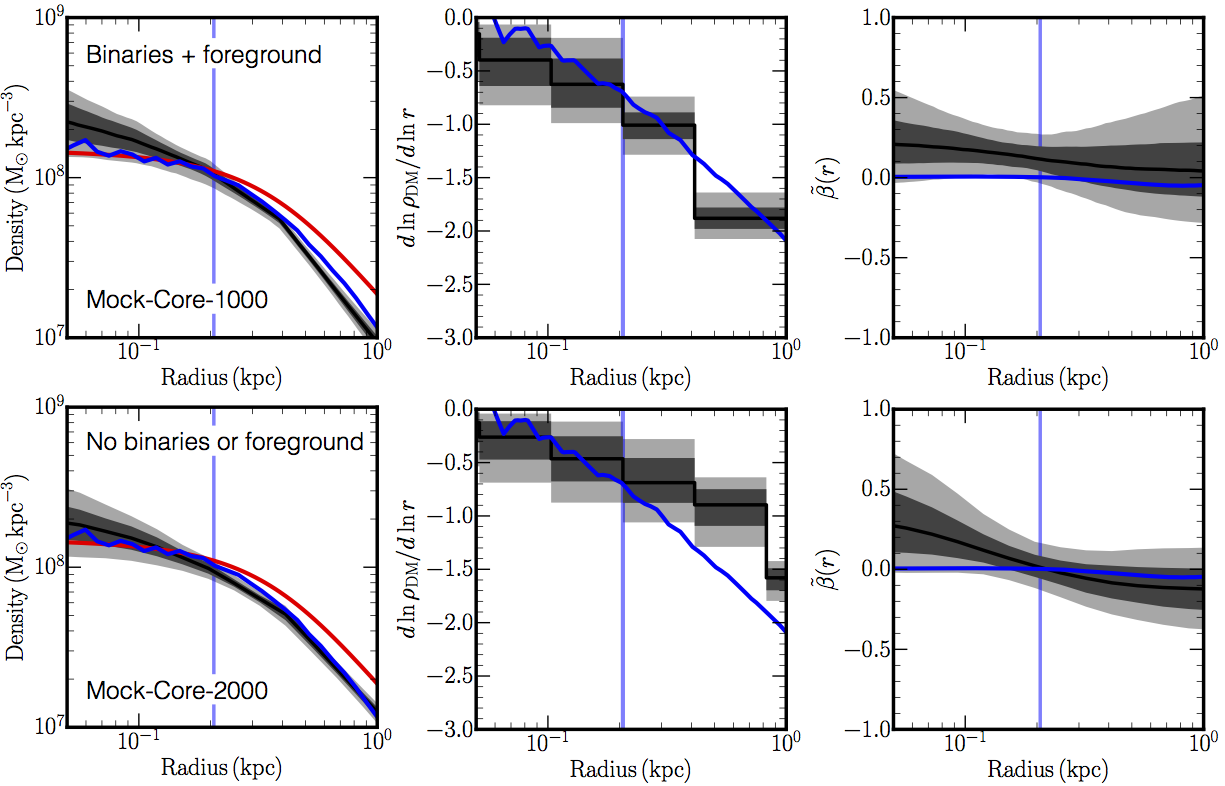}
\caption{As Figure \ref{fig:mock} for the Mock-Core mock, but with 1000 stars with spectroscopic data (top) and with 2000 stars with spectroscopic data but without binary stars or foreground contamination (bottom).}
\label{fig:mock_bias_core}
\end{center}
\end{figure*}

In this Appendix, we explore the effects of binary stars, foreground contamination and the spectroscopic sample size on the Mock-Core mock. The results are shown in Figure \ref{fig:mock_bias_core}, where the lines and contours are as in Figure \ref{fig:mock}. The top row shows how the results change when doubling the spectroscopic sample size to 1000 stars. Now the slight bias towards cuspy models seen in Figure \ref{fig:mock} (middle row) is gone. However, some bias in the recovered velocity anisotropy profile (Figure \ref{fig:mock_bias_core}, too row, right panel) remains. The bottom row shows what happens if we increase the spectroscopic sample size further to 2000 stars and remove the binary stars and foreground contamination. The results for the density profile further improve, while now there is only some weak bias towards radial anisotropy at the centre (Figure \ref{fig:mock_bias_core}, bottom row, right panel). These tests demonstrate that both the uncertainty and bias in the recovery of the dark matter density profile depend primarily on the spectroscopic sample size. Binary stars and foreground contamination induce some small bias, particularly in the velocity anisotropy profile, but their effect is largely benign.

\section{Testing the robustness of our \GravSpherebf\ models for Draco}\label{app:draco_robust}

\begin{figure*}
\begin{center}
\includegraphics[width=0.99\textwidth]{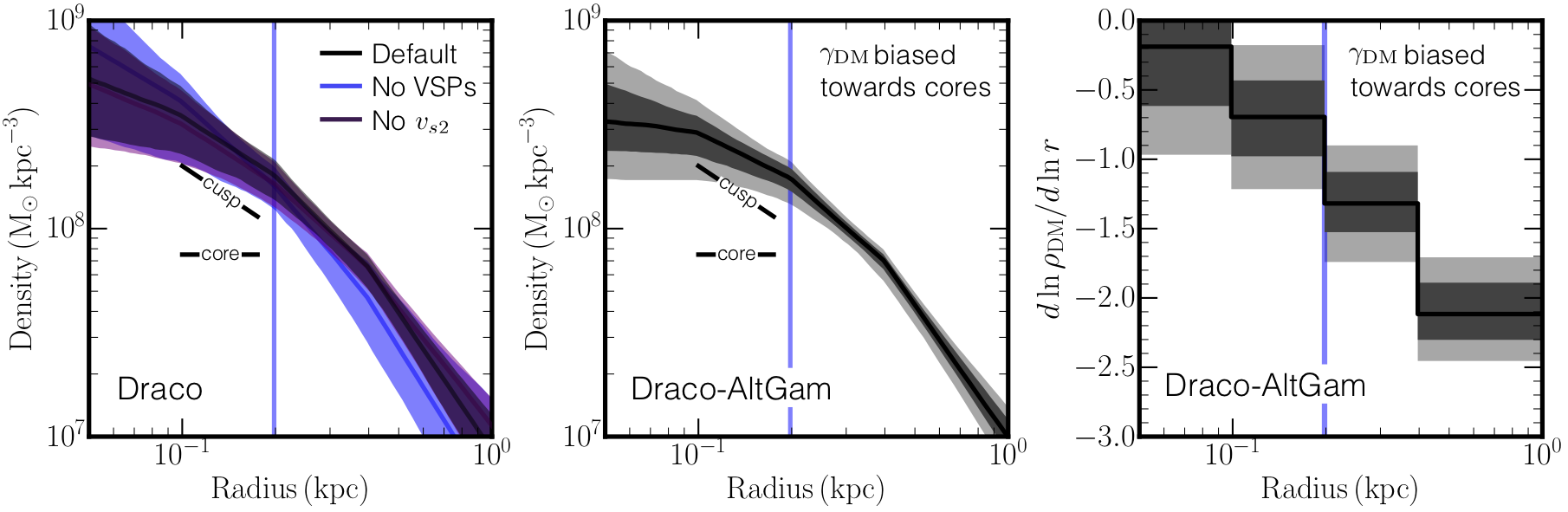}
\caption{Testing the robustness of our \GravSphere\ models for Draco. The left panel shows the 95\% confidence intervals of the dark matter density profile for \GravSphere\ models run without VSPs (blue) and without $\vstwo$ (purple) as compared to our default model with VSPs (black). The lines and contours are as in Figure \ref{fig:draco}. The middle panel shows similar results for Draco with a rather extreme prior on $\gamma_{\rm DM}$ that biases the models towards dark matter cores. The right panel shows $\gamma_{\rm DM}(r) \equiv d\ln\rho_{\rm DM}/d\ln r$ for this same model.}
\label{fig:draco_robust}
\end{center}
\end{figure*}

In this Appendix, we explore the robustness of our \GravSphere\ models for Draco. In Figure \ref{fig:draco_robust} (left panel), we show results for \GravSphere\ models run without VSPs (blue), without $\vstwo$ (purple) and with our default choice of $\vsone+\vstwo$ (black). The contours show the 95\% confidence intervals of the radial dark matter density profile in each case. As can be seen, the VSPs improve the constraints for $R > R_{1/2}$ and, to a lesser extent, for $R < 0.25 R_{1/2}$. However, in all cases Draco favours a large density over the range $0.5 < R/R_{1/2} < 1$, consistent with a CDM cusp.

In Figure \ref{fig:draco_robust}, middle and right panels, we explore the effect of changing our priors on the logarithmic density slope. Our default priors allow a range $-3 < \gamma_{\rm DM} < 0$ in each mass bin. However, as pointed out in \S\ref{sec:mock}, in the absence of sufficiently constraining data, this can lead to a bias towards cuspier models (since models with a flat core, $\gamma_{\rm DM} = 0$, occupy a smaller hypervolume of the parameter space than cuspy models with $\gamma_{\rm DM} = -1$). To test the effect of this bias on our results for Draco, we re-ran our \GravSphere\ model chains assuming a flat prior over the range $-3 < \gamma_{\rm DM}' < 2$, and setting $\gamma_{\rm DM} = 0$ if $\gamma_{\rm DM}' > 0$ and $\gamma_{\rm DM} = \gamma_{\rm DM}'$ otherwise. In the absence of constraining data, this rather extreme prior biases \GravSphere\ towards cores by creating a large region of hypervolume in which $\gamma_{\rm DM} = 0$. As can be seen, this new prior has the effect of making Draco less steep inside $R_{1/2}$, systematically pushing $\gamma_{\rm DM}(R < R_{1/2})$ towards cores (Figure \ref{fig:draco_robust}, right panel). As for our mock data tests in \S\ref{sec:mock}, this shift is smaller than our 68\% confidence intervals on $\gamma_{\rm DM}$, but is nonetheless a source of systematic uncertainty on our recovery of $\gamma_{\rm DM}(R < R_{1/2})$. By contrast, the amplitude of the inner density at 150\,pc is not significantly changed. We find $\rho_{\rm DM}(150\,{\rm pc}) = 2.1_{-0.6}^{+0.5} \times 10^8\,{\rm M}_\odot \,{\rm kpc}^{-3}$, with a logarithmic slope of $\gamma_{\rm DM}(150\,{\rm pc}) = -0.70_{-0.52}^{+0.52}$ at 95\% confidence.

Note, as pointed out in \S\ref{sec:priors}, we consider the above modified prior on $\gamma_{\rm DM}$ to be rather extreme. It removed a small bias on our Mock-Core mock towards cusps while introducing a larger bias on our Mock-Cusp towards cores (Figure \ref{fig:mock_gambias}). Nonetheless, even with this prior, our \GravSphere\ models for Draco favour a high density and steep logarithmic slope at 150\,pc, consistent with a CDM cusp (\S\ref{sec:cuspcore}).

\bibliographystyle{mnras}
\bibliography{final_refs}

\end{document}